\documentclass[12pt]{article}
\usepackage{amsmath,amssymb,amsthm,amsfonts}
\usepackage{wasysym}
\usepackage{graphicx}
\usepackage{subcaption}
\usepackage{xcolor}
\usepackage{stackengine}
\usepackage[colorlinks]{hyperref}
\usepackage{multirow}
\usepackage{algorithm}
\usepackage{algorithmic}

\usepackage{geometry}
\usepackage{appendix}

\theoremstyle{definition}

\newcommand{\ea}{\textit{et al. }}
\newcommand{\del}{\nabla}

\renewcommand{\epsilon}{\varepsilon}

\newcommand{\D}{\mathbf{D}}
\newcommand{\x}{\mathbf{x}}

\begin{document}

\title{Tumor ablation due to inhomogeneous -- anisotropic diffusion in generic 3-dimensional topologies}

\author{Erdi Kara\footnotemark[2],
Aminur Rahman\thanks{Corresponding Author, \url{amin.rahman@ttu.edu}} \thanks{Department of Mathematics and Statistics, Texas Tech University}, Eugenio Aulisa\footnotemark[2],
Souparno Ghosh \thanks{Department of Mathematics, University of Nebraska - Lincoln}}

\date{}
\maketitle

\begin{abstract}
We derive a full 3-dimensional (3-D) model of inhomogeneous -- anisotropic
diffusion in a tumor region coupled to a binary population model.  The diffusion
tensors are acquired using Diffusion Tensor Magnetic Resonance Imaging (DTI) from
a patient diagnosed with glioblastoma multiform (GBM).  Then we numerically simulate
the full model with Finite Element Method (FEM) and produce drug concentration heat
maps, apoptosis regions, and dose-response curves.  Finally, predictions are made about
optimal injection locations and volumes, which are presented in a form that can be
employed by doctors and oncologists.
\end{abstract}



\section{Introduction}
\label{Sec: Intro}

Among the various types of tumors, brain tumors are associated with very high mortality.
The five-year survival rate for people with a malignant brain or central nervous system
tumor is approximately 34\% for men and 36\% for women in the United States and brain
tumors account for 85\% to 90\% of all primary central nervous system tumors
\cite{cancerstat}.

One of the major obstacles to improve current treatments is the presence of some physical barriers such
as the blood-brain barrier and the blood–brain tumor barrier impeding drugs from reaching the tumor sites
in the brain \cite{parodi2019established, wei2014brain, daneman2012blood}. The blood-brain barrier (BBB),
existing between the brain’s microvessels and tissue, prevents many macromolecules from entering interstitial
space in the brain; thus separating the central nervous system and systemic circulation of the body.  It is reported
in \cite{de2007drug} that BBB prevents entry of approximately 98\% of the small molecules and nearly 100\% of
large molecules, such as recombinant proteins or gene-based medicines into brain tumors from the vascular compartment.
Similarly, the blood-brain tumor barrier (BBTB) is located between brain tumor tissues and the brain's blood vessels formed
by highly specialized endothelial cells, limiting the delivery of most anti-cancer drugs to tumor tissue. To overcome the
challenges associated with these barriers, several methods have been developed such as intrarterial administration, barrier
disruption, drug packaging, and inhibiting drug efflux from tumors \cite{groothuis2000blood}.  Regardless, oral and
intravascular administration allows only a small fraction of therapeutic agents to reach the tumor region in the brain
because the drug concentration decreases precipitously due to the sink effect of the extracellular space along the route
of drug transport to the tumor region.  Thus, necessitating the administration of a high dose to achieve sufficient drug
concentrations to kill the tumor cells.  Unfortunately, physiological toxicity limits the amount of therapeutic agent allowed in a
particular therapy.

Significant effort has gone into developing therapies with high efficacy to toxicity ratios through the use
of lower drug concentrations via direct administration in targeted regions.  One such method is discussed in
de Boer \ea \cite{de2007drug}.  Drug injection therapies allow for uncomplicated individualized treatment for solid
accessible tumors \cite{de2007drug, CSRYL13, BMS14, Morhard2017, SCLAMGL18}.  Further, these types
of therapies provide us with an opportunity to optimize drug efficacy by changing the
fluidic properties of the injection as shown by Morhard \ea \cite{Morhard2017}.  While
testing a variety of therapeutic agents on animals is time intensive
and costly, numerical simulations may prove to be a cheap and effective solution.

In order to conduct numerical experiments on the effects of drugs on cancerous tumors,
we need transport -- population coupled models.  This may seem counter-intuitive to
the goal of developing fast methods to determine efficacy and toxicity due to how
computationally intensive sophisticated coupled models can be.  Recent review articles
and books show evidence of this in the wealth of transport models
\cite{WaiteRoth12, KGR2013}, and separately in the variety of population models
\cite{KNE2016}; yet there is a dearth of coupled transport -- population models.
Further, while there are many sophisticated transport models for drugs penetrating
into the tumor from the blood stream (e.g. \cite{SoltaniChen2011, SoltaniChen2012, SSRSBBM15}),
there are few mathematical investigations of drugs injected into the tumor.
 
In \cite{RGP18}, Rahman \ea present a simple drug diffusion -- binary population
model.  It is assumed that a drug is being injected directly into the center of a
homogeneous -- isotropic spherical tumor, and hence the diffusion is radial with
constant diffusivity.  Furthermore, it is assumed that a cell is either dead if
the concentration of the drug is greater than a threshold and alive if the concentration
is less than this threshold; i.e., the drug acts as a trigger for cell death.  From
this model dose-response curves (response as a function of dose) are produced in order
to relate it to relevant empirical data, such as Harvard Medical School's LINCS data
set.  Since the data set does not include replication studies, artificial replication
dose-response curves were produced, and the dose-response curves from the mechanistic
model were shown, in many cases, to lie within 95\% piecewise-linear confidence bands.

While the model of Rahman \ea \cite{RGP18} performed well against artificial replication
data, the simplicity is burdened by the baggage of assumptions.  Transport of drugs in
the brain is a complex process due to the highly inhomogeneous and anisotropic
structure of brain tissue, local pressure differences, and chemcial interactions of the
drug with the surrounding tissue. However, a model that
does not obviate negligible contributions falls prey to computational constraints.
Even with parallelization on a supercomputing cluster, would it be useful to an
oncologist that is working directly with a patient?  

In this investigation, we keep the simple binary population model, and explore the complexities
of drug transport from a single injection into deformed globular tumors (topological $3$-spheres).
It is well known that diffusion dominates, but we assume it can be inhomogeneous and anisotropic.
This allows for fast \textit{in vivo} simulations of direct injection therapies, which we use to
produce dose-response curves, apoptosis regions, and optimal injection locations.  This investigation
endeavors to present results to aid practitioners in optimizing treatment strategies.

The remainder of the paper is organized as follows:  We begin our discussions by deriving
the inhomogeneous - anisotropic diffusion and binary population models in Sec. \ref{Sec: GenericDiffusion}.
Then in Sec. \ref{Sec: Numerics}, we develop the computational foundation of the investigation
in two steps.  First, Sec. \ref{Sec: FEM Discretization} sets up the finite element scheme.  We use
Galerkin Finite Element Method for the spatial discretization and a Crank-Nicolson scheme for
the temporal integration.  Then the diffusion tensors are constructed from patient MRI data
in Sec. \ref{Sec: DTI}.  Section \ref{Sec: Results} presents the numerical simulations and
oncological predictions of our study.  We first examine the drug diffusion in the tumor and
observe the high inhomogeneity and anisotropy in the concentration heat maps.  Then the
apoptosis is simulated by invoking the binary population model.  This allows us to create
dose-response curves.  Importantly, the model predicts optimal injection sites, evidenced by
the dose-response curves, different from what intuition might suggest.  Finally, the investigation
is concluded in Sec. \ref{Sec: Conclusion} with a discussion on viable oncological applications and
future modeling directions.

\section{Drug diffusion and binary population models}\label{Sec: GenericDiffusion}

In this section we derive the inhomogenous -- anisotropic diffusion model and the
binary tumor population model.  First we write a general diffusion model.
It has been shown through Magnetic Resonance Imaging (MRI) in the brain that tumors
often grow in an inhomogeneous -- anisotropic diffusion-like manner
\cite{KCMSWMDA10, MLHKWAG11, CGFBAC15, EKS16}.  
Using Diffusion Tensor Imaging (DTI) techniques \cite{ODonnell-Westin2011}, the
effective diffusivity of water in tissue can be estimated. Since there is a large contrast between
cancerous and healthy tissues, DTI can be used to map the geometry of a tumor and estimate the
diffusivity of water in a tumor \cite{RMSZKM12, MekkaouiDTI16, AEDS18}.

Consider a drug with molarity $u$ diffusing from an injection into a porous tumor, with an effective
diffusivity tensor $\D$. Next for the sake of brevity, let $\x = (x_1,x_2,x_3)$ be the position vector.
Finally, we model the leak at the boundary of the region of interest as ``Newton's law of cooling''
\cite{NewtonCooling}.  This gives us the generic model in Cartesian coordinates
\begin{subequations}
\label{Eq: GenericModel}
\begin{align}
&\frac{\partial u(\x;t)}{\partial t} = \del \cdot \left(\D(\x)\del u(\x;t)\right),\quad
\x \in \Omega;\label{Eq: GenericDiffusion}\\
&\D(\x)\del u(\x;t) \cdot n = -\gamma u(\x;t), \quad \x \in  \partial\Omega;
\label{Eq: GenericNewtonCooling}\\
&u(\x; t = 0) = \bar{u}(\x),\label{Eq: GenericIC}
\end{align}
\end{subequations}
where $\bar{u}(\x)$ is some generic initial concentration
profile of the drug soon after injection and $\gamma$ is the constant leak coefficient.
Since all the cells at the boundary are assumed to be non-cancerous, 
the rate of leakage is equivalent to the rate of diffusion at the boundary.

We now impose a specific initial condition.  In \cite{RGP18} a bump function (compact
Gaussian) decaying to zero just within the domain was used.  During
inhomogeneous -- anisotropic diffusion, the drug does not diffuse evenly, and hence
a bump function that extends to the endpoints would not capture the irregularities
expected in such a problem.  Nevertheless, there are many numerical advantageous to
using a bump function.  We may either introduce a sharper bump or try to capture the
diffusive irregularities in the initial condition.  In general, for an injection at point $\x_c$, we have
\begin{equation}
\label{Eq: IC}
\bar{u}(r)= \dfrac{U_{0}}{V_{b}}
\begin{cases} 
      \text{exp}\left(1-\dfrac{R_{b}^2}{R_{b}^{2}-r^{2}}\right), & r < R_{b} \\
      0, & r \geq R_{b}
  \end{cases}
\end{equation}
where, $r = \left\|\mathbf{x-x_{c}}\right\|$, $R_\text{b}$ is the radius of the bump, $U_0$ is the injected concentration,
and $V_\text{b}$ is the volume of the bump function in spherical coordinates; that is,
\begin{align*}
V_\text{b} &= \int_0^\pi\int_0^{2\pi}\int_0^{R_\text{b}}
\exp\left(1 - \frac{R_\text{b}^2}{R_\text{b}^2-r^{2}}\right)r^2\sin\phi dr d\theta d\phi
\end{align*}

\subsection{Binary population model}
\label{Sec: Binary Population}

As done in \cite{RGP18}, we use a binary population model: the tumor cell is dead after some exposure time
$\tau$ (which is much larger than the diffusive time scale), if at any time during the diffusion process the drug
concentration is above some given threshold value $u_T(\tau)$, otherwise it is alive.  To simplify computation,
the model assumes natural cell death rate is equivalent to the cell population growth rate, and effectively negate
each other.  We expect the threshold to decrease with time because it takes more toxins to kill a cell quickly than
it does to kill it slowly.  Further, since empirical studies often plot response data against log time, we expect
$u_T$ to be a negative exponential with time $\tau$,
\begin{equation}
u_T(\tau) = a - be^{-c\tau},
\label{Eq: Threshold}
\end{equation}
where the parameters $a$, $b$, and $c$ are back calculated from a representative
sample of the Harvard Medical School LINCS drug data set.  From the representative
sample we calculate what $u_T(\tau = 24 \text{ hrs})$, $u_T(\tau = 48 \text{ hrs})$,
$u_T(\tau = 72 \text{ hrs})$ must be in order to produce the empirically observed
response.  Then we have three equations with three unknowns, which is solved
explicitly in \cite{RGP18}. In this paper, we use the following threshold values.

\begin{equation*}
u_{T}(24)=0.230153 \text{ $\mu$M}, \quad u_{T}(48)=0.0700055 \text{ $\mu$M},
\quad u_{T}(72)=0.0499662 \text{ $\mu$M}
\end{equation*} 


\section{Numerical procedure}\label{Sec: Numerics}


In this section, we first describe the numerical procedure to approximate the
solution of the system \eqref{Eq: GenericModel} in Sec. \ref{Sec: GenericDiffusion}.
Galerkin finite element method (GFEM) is used for the spatial discretization and a
Crank-Nicolson scheme for the temporal integration. For a practical introduction
finite element method, reader can refer to \cite{larson2013finite}.

For the inhomogeneous -- anisotropic diffusivity we incorporate diffusion tensors
from diffusion tensor magnetic resonance imaging (DTI) data.
The magnetic field gradients in different directions from the MRI is used to map
out directions of faster and slower diffusion that is normalized to the diffusivity
of water molecules.  While in medical imaging the diffusion tensor is predominantly
used to identify anomalies, and often averaged out in preferential directions for
that reason, we use it in the diffusivity of the transport model.
Since the only errors are from the DTI data and GFEM, we can produce accurate
qualitative simulations of \textit{in vivo} scenarios.

There are a few studies in the literature integrating FEM with DT-imaging in a modeling framework. 
Kraft \textit{et al.} incorporated DTI with FEM to investigate the mechanics of neurotrauma
\cite{kraft2012combining}. In \cite{reslimb}, Ramasamy \textit{et al.} proposed a subject-specific
finite element model for the residual limb to assess the effect of a particular socket on deep tissue
injury. They utilized DTI to reveal the anatomy of muscle fiber and mapped the information onto a
finite element mesh.  A nonlinear hyperelastic, transversely isotropic skeletal muscle constitutive
law containing a deep tissue injury model were then solved with FEM. Clatz \textit{et al.} \cite{clatz2005realistic}
used FEM to simulate the invasion of GBM in the brain parenchyma and its mass effect on the invaded
tissue.  They described a coupling strategy between reaction-diffusion and linear elastic mechanical
constitutive equations where diffusion tensor information was provided by DTI.  In \cite{Stoverud2012},
a convection-enhanced drug delivery (CED), where the anti-cancer agent is directly administered into the
brain tissue, was introduced. Governing equations concerning the transport of the therapeutic agent and
tissue deformation was solved with the Finite Volume Method, where the information about the structures of
the tissue is acquired through DTI.

\subsection{Finite element discretization }\label{Sec: FEM Discretization}

Consider the Sobolev space $H^{1}(\Omega)=\{v \in L^{2}(\Omega): \partial_{x_{i}} v
\in L^{2}(\Omega)\quad i=1,2,3 \}$, where $\Omega$ is the domain of the PDE from Sec.
\ref{Sec: GenericDiffusion}. If $u$ is sufficiently smooth, by multiplying \eqref{Eq: GenericDiffusion} with a test function $v \in H^{1}(\Omega)$ and integrating over $\Omega$ using Green's formula, we
obtain the variational formulation of \eqref{Eq: GenericDiffusion}; that is, find
$u$ such that for every $t \in I=[0,T]$, 
\begin{equation}
\label{Eq: WeakForm}
\int_{\Omega} \frac{\partial u}{\partial t}vdx=-\int_{\Omega}\mathbf{D}\nabla u\cdot\nabla v dx -\int_{\partial\Omega} \gamma uv ds, \hspace{0.5cm} \forall v \in H^{1}(\Omega), \quad t \in I.
\end{equation}
Let
\begin{equation}
\label{eq:tetra}
\Omega_{h}=\{K_{1},K_{2},..,K_{n} \}
\end{equation}
be a geometrically conforming hexagonal triangulation of $\Omega$. As a test space, we use the space of scalar valued piece-wise quadratic polynomials; i.e.,
\begin{equation*}
V_{h}=\{p \in C^{0}(\bar{\Omega})\ :\: p|_{K} \in Q_{2}(K), \hspace{0.5cm} \forall K
\in \Omega_{h} \} \subset H^{1}(\Omega).
\end{equation*}
Replacing $H^{1}(\Omega)$ with $V_{h}$, the finite element formulation of
\eqref{Eq: WeakForm} reads: find $u_{h}$ such that for every
$t \in I$,
\begin{equation}
\label{Eq: FEM form}
\int_{\Omega} \frac{\partial u_{h}}{\partial t}v_{h}dx=-\int_{\Omega}\mathbf{D}\nabla u_{h}\cdot\nabla v_{h}dx -\int_{\partial\Omega} \gamma u_{h}v_{h}ds, \hspace{0.5cm} \forall v_{h} \in V_{h}, \quad t \in I.
\end{equation}

Let $\{ \psi_{i}\}_{i=1}^{n}$ be a basis of the space $V_{h}$ consisting of the
orthogonal nodal basis functions satisfying $\psi_{i}(N_{j})=\delta_{ij}$ for
every nodal point $N_{j}$. For every $u_{h} \in V_{h}$, there exist time dependent
coefficients $\xi_{j}(t)$ such that 
\begin{equation}
\label{Eq: LinearCombination}
u_{h}(x,t)=\sum_{j=1}^{n} \xi_{j}(t)\psi_{j}(x).
\end{equation}
From this construction,  $\xi_{j}(t)$'s are the nodal values of $u_{h}$ for every
time $t$. If we substitute \eqref{Eq: LinearCombination} in \eqref{Eq: FEM form},
we obtain the following system of ordinary differential equations (ODEs)
\begin{equation}
\label{Eq: OdeSys}
M\dot{\xi(t)}=-(A+R)\xi(t),
\end{equation}
where $\xi(t)=[\xi_{1}(t),\xi_{2}(t),..,\xi_{n}(t)]^{T}$ and
\begin{equation}
\label{eq:matequ}
M_{ij}=\int_{\Omega} \psi_{j}\psi_{i}, \quad A_{ij}=\int_{\Omega}\mathbf{D}\nabla\psi_{j}\cdot\nabla\psi_{i}, \quad R_{ij}=\int_{\partial\Omega}\gamma \psi_{j}\psi_{i}.
\end{equation}

Note that the diffusion tensor $\mathbf{D}$ is defined element-wise. Let
$0=t_{0} <t_{1}< \ldots <t_{N}=T$ be a partition of the interval $[0,T]$
with the constant time step $\Delta t=t_{n+1}-t_{n}$. Application of the
Crank-Nicolson scheme for \eqref{Eq: OdeSys} yields
\begin{equation}
\label{Eq: BackEuler}
M\frac{\xi_{k+1}-\xi_{k}}{\Delta t}=-(A+R)\frac{\xi_{k}+\xi_{k+1}}{2},
\end{equation}
where $\xi_{0}$ is chosen as the nodal values of $\bar{u}(x)$  defined in
\eqref{Eq: GenericIC} of Sec. \ref{Sec: GenericDiffusion}. For the remainder
of the manuscript we take the step size $\Delta t=0.02$ with $N=60$ equal
time intervals and the leak coefficient is set to $\gamma=0.002$. For the implementation of the problem \eqref{Eq: BackEuler}, we use the open source finite element C++ library FeMUS \cite{femus}.

\subsection{Incorporating the diffusion tensor}\label{Sec: DTI}

Diffusion patterns of water molecules in biological tissue can be visualized by
means of diffusion tensor magnetic resonance imaging (DTI). The diffusivity in the medium is quantified
at each image voxel (a volumetric pixel) with a diffusion tensor that relates
diffusive flux to a concentration gradient in each Cartesian direction. The three
diagonal elements $D_{xx}$, $D_{yy}$, and $D_{zz}$ represent diffusion coefficients
measured along each of the principal $x-$, $y-$, and $z-$ axes. The six off-diagonal
entries quantify the correlation of Brownian motion between corresponding principal
directions. A generic diffusion tensor can be written as 
\begin{equation}
\mathbf{D}=
\begin{bmatrix}
 D_{xx} & D_{xy} & D_{xz} \\
 D_{yx} & D_{yy} & D_{yz} \\
 D_{zx} & D_{zy} & D_{zz}
\end{bmatrix},
\label{DTMatrix1}
\end{equation}
where $D_{ij}$ has the unit of $mm^{2}/sec$.

The diffusivities from DTI are usually averaged out and illustrated by
condensing the tensor information into a scalar quantity or plotted as a color
encoded texture map. The former consists of scalar measurements to quantify
the magnitude or the shape of the diffusion. In terms of magnitude, mean
diffusivity (MD), which is the mean of the eigenvalues of the diffusion
tensor, is one of the most common scalar measurements. On the other hand,
fractional anisotropy (FA), which is the normalized variance of the eigenvalues, is
the most commonly used anisotropy measure. In addition to various scalar measurements,
one can also consider the direction of the major eigenvector (the eigenvector associated with
the largest eigenvalue) and create a color
map for the corresponding directions. The most commonly used color scheme in
terms of anatomical planes is as follows; blue is superior-inferior, red is
left-right, and green is anterior-posterior \cite{ODonnell-Westin2011}. For
an extensive overview of the diffusion tensor imaging, the reader may refer
to \cite{DtiBook}.

\begin{figure}[htbp]
\centering
\stackinset{l}{0mm}{t}{2mm}{\color{white}\textbf{\large (a)}}{\stackinset{l}{1.95in}{t}{2mm}{\color{white}\textbf{\large (b)}}{\stackinset{l}{3.9in}{t}{2mm}{\color{white}\textbf{\large (c)}}{\includegraphics[width=0.9\textwidth]{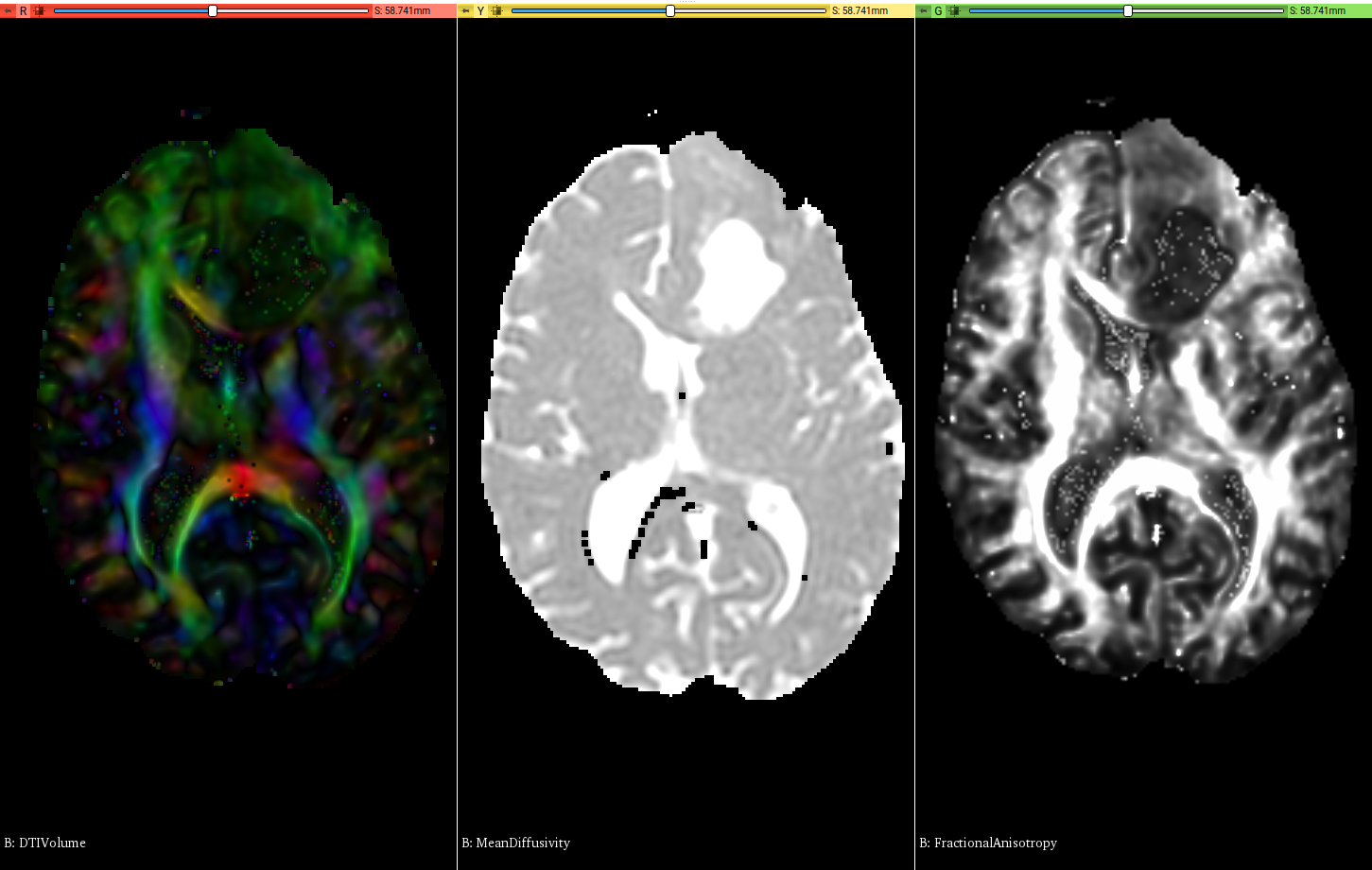}}}}
\caption{Axial view of the brain DTI with illustrations from \textit{3DSlicer} of \textbf{(a)} the major eigenvalue direction,
\textbf{(b)} fractional anisotropy, and \textbf{(c)} mean diffusivity.  In all images the tumor region
can be seen around the right frontal lobe.}
\label{fig:TumorAxial}
\end{figure}

To capture the anisotropies of the diffusion in our numerical simulations we use a
dataset that includes a diffusion tensor magnetic resonance image of a 35-year old male diagnosed with
glioblastoma multiform (GBM). The dataset can be found in the tutorial \cite{DtiData}.
From the diffusivity data, in Fig. \ref{fig:TumorAxial}, we derive and display the major
eigenvector direction (indicated by the colors prescribed in the preceding paragraph),
fractional anisotropy, and mean diffusivity.  In the figure the GBM region can be observed
around the right frontal lobe.

The coordinate dimensions of the sample volume is $256 \times 256 \times 51$
with voxel size $1 \times 1 \times 2.6$ $\text{mm}$. In the pre-processing of
the DTI volume, we utilized the open source software \textit{3DSlicer} \cite{3dslicer},
which allows advanced medical image analysis and processing. It provides a 
graphical user interface with various modules as well as a Python console,
which gives access to data arrays of image models for further analysis.

The diffusion tensor is symmetric and positive definite (SPD) but in practice the
positive definiteness can be corrupted due to measurement noise. Thus we first
re-sample the DTI volume to correct the tensors that are not positive semi-definite.
We are particularly interested in diffusion in the tumor tissue. Hence, we extract
a region of $64 \times 64 \times 24$ voxel-wise enclosing the entire tumor region from
the re-sampled volume and scale it into a cube $\Omega$ of sides $5$ centered at the
origin,
\begin{equation}
\label{eq:domain}
\Omega=\{ (x,y,z) : -2.5\leq x,y,z\leq 2.5\}.
\end{equation}
We take $\Omega$ as the computational domain of the problem described in
(\ref{Eq: GenericModel}). After the re-scaling, the tumor region fits completely
inside the rectangular subdomain of $\Omega$,
\begin{equation}
\label{eq:tregion}
\mathbb{T}=\{ (x,y,z) : -0.6\leq x\leq 1.7, -1.5\leq y \leq 1.5, -1.1\leq z \leq 2.2\} \subset \Omega.
\end{equation}

It is convenient to rescale the tensor $\mathbf{D}$ over the region
$\mathbb{T}$ defined in \eqref{eq:tregion}. Let 
\begin{equation}
\mathbf{D}_{i}=
\begin{bmatrix}
 D_{xx} & D_{xy} & D_{xz} \\
 D_{yx} & D_{yy} & D_{yz} \\
 D_{zx} & D_{zy} & D_{zz}
\end{bmatrix}       
\label{DTMatrix}
\end{equation}
be the diffusion tensor defined for the hexagonal element $K_{i}$ in
\eqref{eq:tetra}. Note that we define $\mathbf{D}$ element-wise so it is
essentially piece-wise discontinuous across the problem domain. Let
$\mathbf{D}_i^a=(D_{xx}+D_{yy}+D_{zz})/3$ be the element-wise apparent
diffusion coefficient (ADC).  We will use Algorithm \ref{alg:TumAlg} to
scale the diffusion tensors in $\Omega$ and also identify the tumor elements
in $\mathbb{T}$.
 
\begin{algorithm}
\caption{}
\begin{algorithmic}
\algsetup{linenosize=\small}
  \scriptsize
\label{alg:TumAlg}
\STATE $N$ = number of elements
\STATE $D$ = diffusion tensor
\STATE $\mathbb{T}$ = region enclosing the tumor bulk
\STATE $AvgTr=0, Counter=0 $
\FOR{$i=1:N$}
\IF{$K_{i}\in \mathbb{T}$}
\STATE $TrE=D_{a}^{i}$
\IF{$TrE>0.002$}
\STATE{$AvgTr=AvgTr+TrE$}
\STATE{$Counter=Counter+1$}
\ENDIF
\ENDIF
\ENDFOR
\STATE {$Scale =AvgTr/Counter$ }
\FOR{$i=1:N$}
\STATE{$D_{i}=D_{i}/Scale$}
\ENDFOR
\end{algorithmic}
\end{algorithm}
 
We set a tumor boundary threshold value of $0.002$ by visual inspection. If our element-wise ADC
$\D_i^a > 0.002$ in the region $\mathbb{T}$ for the raw data, we label it as cancerous
element. With the algorithm above, the mean of ADCs across the assumed tumor elements is scaled
to unity. 

A major prediction made in this investigation is the fraction of the tumor
volume killed against preassigned thresholds, $u_T(24),u_T(48)$ and $u_T(72)$,
from Sec. \ref{Sec: Binary Population}. This task essentially requires the elementwise
identification of the tumor cells in $\Omega$. The only quantitative information
about the sample data are the diffusion tensors at each voxel provided by DTI.
No method is known to precisely differentiate the tumor and healthy cells
by means of diffusion tensor information. However, we employ the fact that
water diffuses significantly faster in GBM tissue than the surrounding healthy
tissue \cite{maier2010diffusion}.

We first assume that there is no tumor cell outside the region $\mathbb{T}$. So, after 
the re-scaling of the diffusion tensor with Algorithm \ref{alg:TumAlg}, we will mark any
element $K_{i}$ in $\mathbb{T}$ as cancerous if $\D_i^a > 0.6$. This is the value where
we observe a relatively sharp transition between the normal and cancerous regions.
To calculate the apoptosis fraction, first consider the unit step function 
\begin{equation}
\label{eq:IC}
H(s)=
\begin{cases} 
      1, & s > 0, \\
      0, & s \leq 0;
  \end{cases}
\end{equation}
and the sub-region $\mathbb{T}_{C} \subset \mathbb{T}$ defined as  
\begin{equation}
\label{eq:subtum}
\mathbb{T}_{C} =  \{ \mathbf{x} \in \mathbb{T}: \mathbf{D}_i^a(\mathbf{x})>0.6 \}
\end{equation}
We compute the fraction of cells, $\beth$, that were once exposed to a concentration higher than $u_T$
at the simulation time $t=t_{n}$ as follows,
\begin{equation}
\beth(\tau,t_n)= \frac{\int_{\mathbb{T}_{C}} H(\max_{t \in[0,t_{n}]}\left[u (\mathbf{x},t) - u_T(\tau)\right])dV}{\int_{\mathbb{T}_{C}} dV}
\label{Eq: Beth}
\end{equation}
where $\tau = 24,\, 48 $ or $72$ hours.  Then to get the apoptosis fraction, $\aleph$,
we prescribe $t_n = T$ in \eqref{Eq: Beth},
\begin{equation}
\aleph(\tau) = \beth(\tau,T).
\label{Eq: Aleph}
\end{equation}
Note that the resulting fractions above are relative
to the sub-region $\mathbb{T}_{C}$ since we assume all tumor cells lie in $\mathbb{T}$.

\begin{figure}[htbp]
\centering
\includegraphics[width=0.9\textwidth]{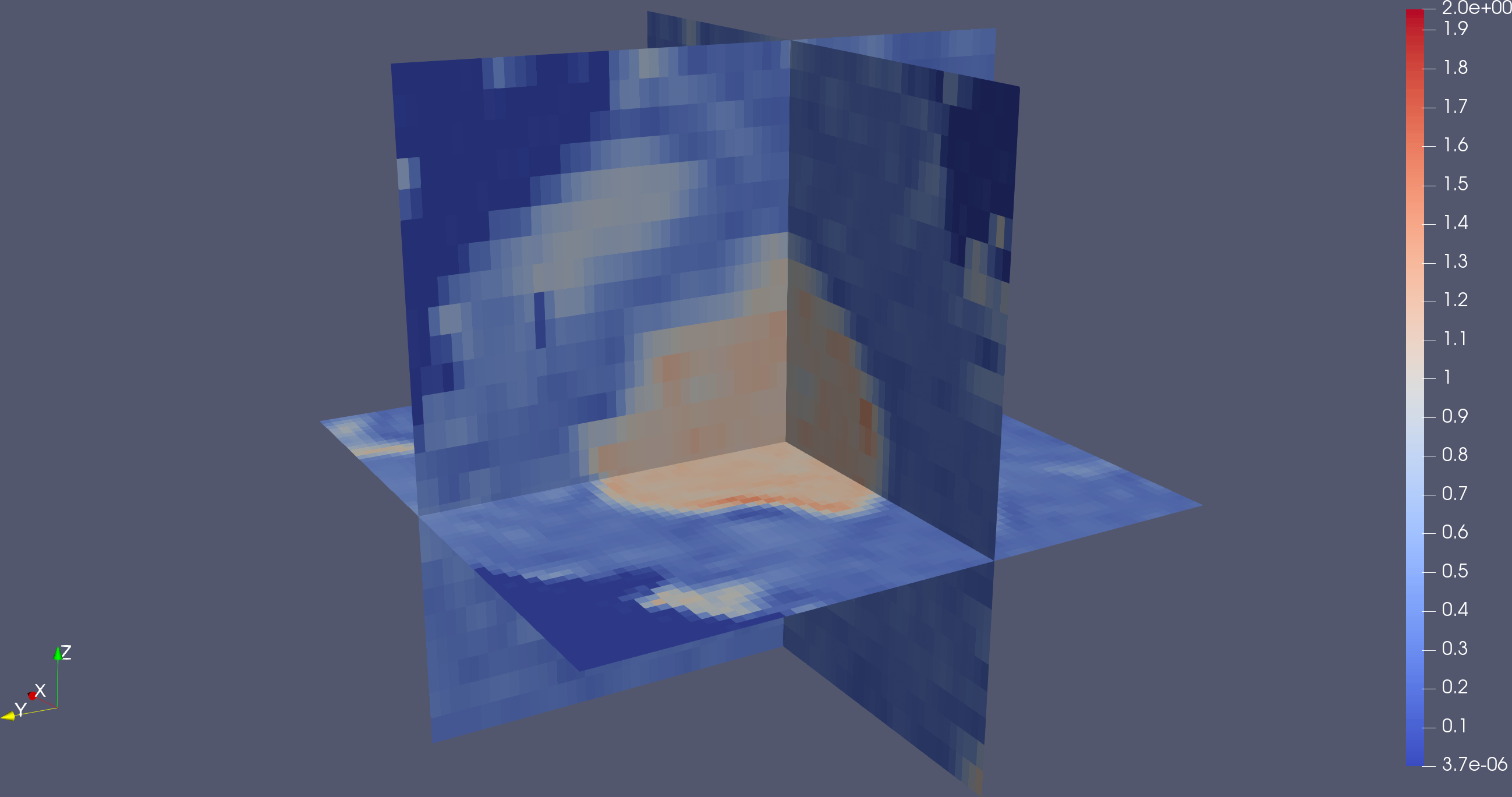}
\caption{Computational domain with three representative slices centered at $\x_c = (0.5,0,0)$.
The heat map indicates the magnitude of the apparent diffusion coefficient (ADC) after the scaling
of the raw diffusion tensor data with Algorithm \ref{alg:TumAlg}.}
\label{fig:TumorRegion}
\end{figure}

Different cross sections of the computational domain $\Omega$ indicating the element-wise
ADCs can be seen in Fig. \ref{fig:TumorRegion}. We construct $\Omega$ to be consistent with
the anatomical coordinate system described in \cite{anacoord}. Positive directions of the $x$, $y$,
and $z$ axes are chosen to be anterior, left, and superior, respectively. By this way, $xy$, $xz$,
and $xz$ coordinate planes correspond to transverse, frontal, and sagittal planes, respectively. In
Fig. \ref{fig:TumorRegion}, for example, the horizontal plane with respect to the monitor corresponds
to the transverse plane.

DTI is a non-invasive technique based on the measurement of the diffusion of water molecules.
Therefore, the diffusion tensor $\mathbf{D}$ may be quantitatively different across the tissue
of interest when another substance, such as a therapeutic agent, is used. Seemingly, there is no
experimental study proposing a numerical relationship between the diffusion tensor of water and
other fluids in brain tissue. Therefore, we will assume that although water and corresponding drug
molecules have different fluidic properties, they display qualitatively similar behavior in the same
medium.  With this assumption, we will treat the diffusion tensors that we extracted from the dataset,
described above, as the diffusion tensors of the agent used in the simulated treatment. For information
on treatments of this type, the reader can refer to \cite{clatz2005realistic,Stoverud2012}.

\section{Results and predictions}\label{Sec: Results}

\begin{figure}[hbp]
\centering
\includegraphics[width=0.9\textwidth]{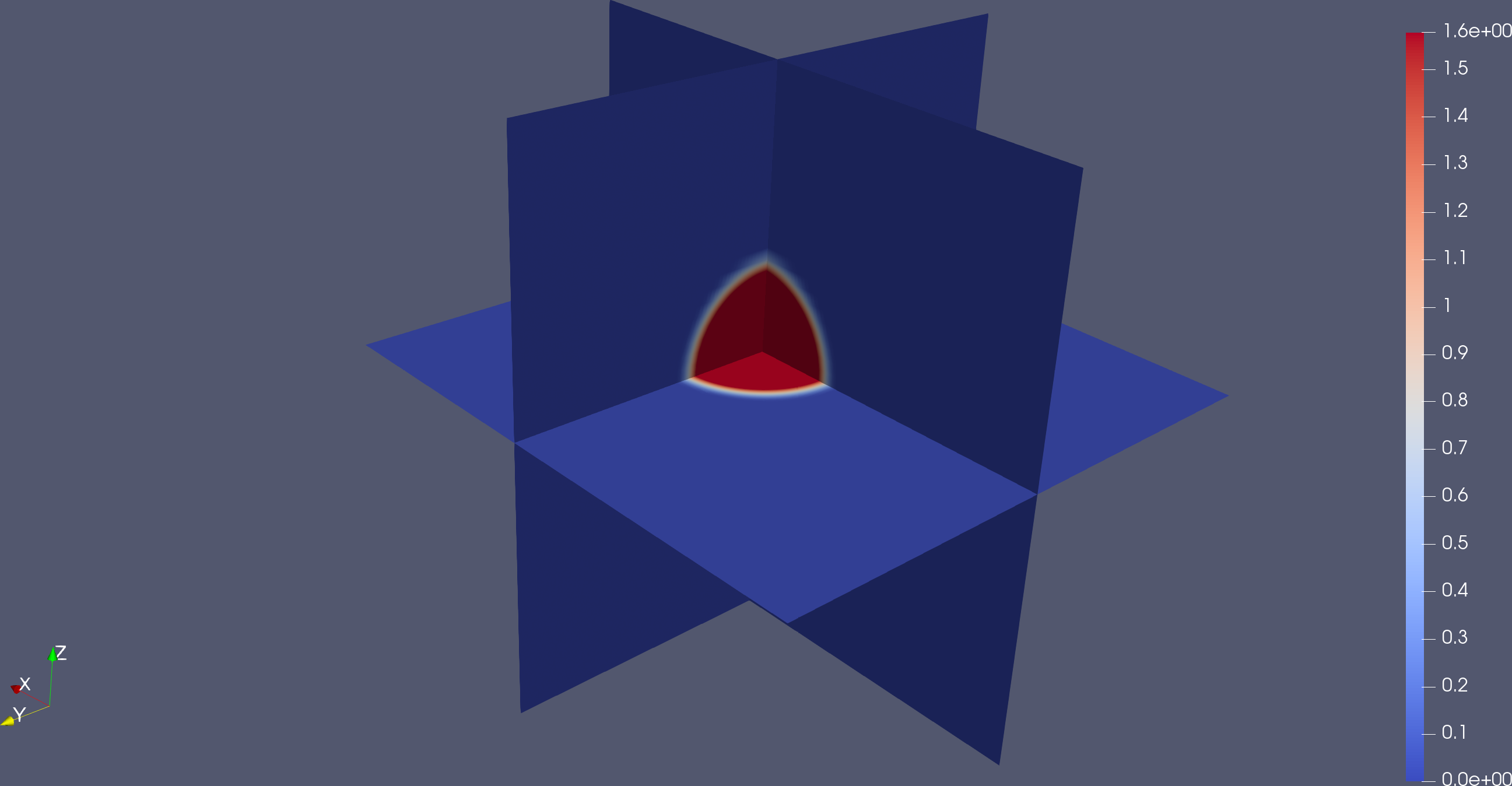}
\caption{Initial concentration profile $u(\x;t=0)$ centered at the location $\x_c = (0.5,0,0)$
with an initial injection molarity of  $U_{0}=1.5$ $\mu\text{M}$.}
\label{fig:bump}
\end{figure}

In this section we present the \textit{in vivo} direct injection treatment simulations resulting from the
numerical solutions of our model \eqref{Eq: GenericModel}.  Based on the location of $\mathbb{T}$, we
set the initial condition (\ref{eq:IC}) (illustrated in Fig. \ref{fig:bump}) as follows
\begin{equation}
\label{eq:initpro}
u(\x;t = 0) = \frac{U_0}{V_\text{b}}\begin{cases}
\exp\left(1 - \frac{1}{1 - r^2}\right) & \text{for $r < 1$},\\
0 & \text{for $r \geq 1$};
\end{cases}
\end{equation}
where $V_{b}=1.1990$ and $r =\sqrt{(x-0.5)^{2}+y^{2}+z^{2}}$.  By Algorithm \ref{alg:TumAlg}, we
scale the patient's water diffusivity tensor to $\D/0.00271734$, which effectively reduces the temporal time
step for computation, and afterwards set the leak coefficient to be $\gamma = 0.002$ $\text{m}/\text{s}$,
which mimics the rate of diffusion for the scaled tensor.  Further, a very high initial concentration of
$U_{0}=1.5$ $\mu\text{M}$ is taken for illustrative purposes.

In Sec \ref{Sec: Binary Population}, we assumed that if the concentration at the element $K_{i}$ is above a certain threshold,
$u_T(\tau)$, at any simulation time, t, then $K_{i}$ will die out after the corresponding exposure times, $\tau$. With
this assumption, we can create a heat map indicating the regions which is predicted to die after the related exposure
times, $\tau = 24$, $48$, and $72$ hours. It should be noted that although the precise tumor region is
$\mathbb{T}_{C}$, we will display the apoptotic region within $\mathbb{T}$ for illustration purposes. 
Also note that the drugs eventually leak out beyond the region $\mathbb{T}_{C}$,
and contribute to toxicity.  Since the response of the healthy cells to the drug is a highly complex phenomenon,
in the figures we will assume that any tissue element experiencing concentration above the thresholds in
Sec. \ref{Sec: Binary Population} will be killed irrespective of the location.

\begin{figure}[htbp]
\centering
\stackinset{l}{0mm}{t}{1mm}{{\color{white}\textbf{\large (a)}}}{\stackinset{l}{2.95in}{t}{1mm}{\color{white}\textbf{\large (b)}}{\includegraphics[width=0.9\textwidth]{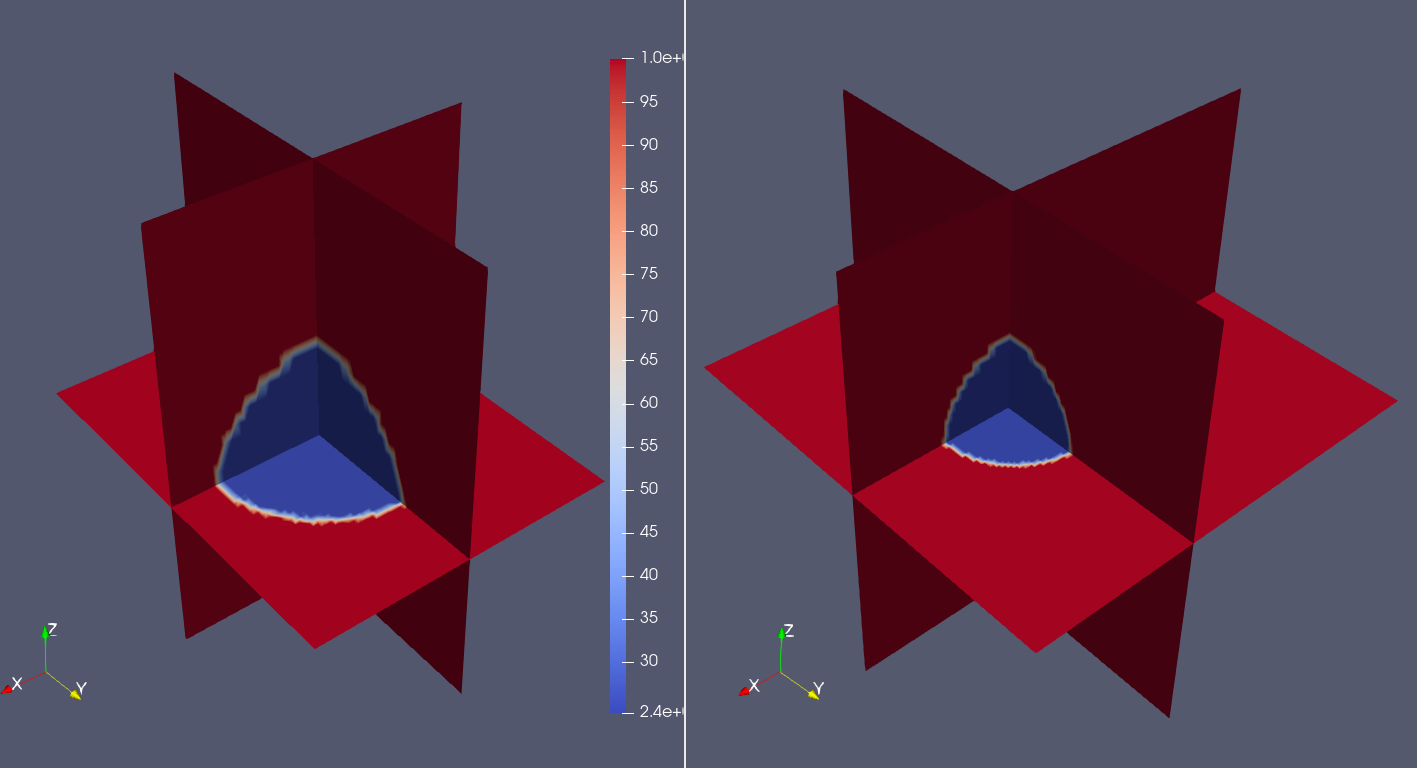}}}
\caption{Apoptotic region induced by the initial bump function centered at $\x_c = (0.5,0,0)$,
with a very high drug concentration of
$U_{0}=1.5$ $\mu\text{M}$ for illustration purposes, within \textbf{(a)} the tumor region $\mathbb{T}$
and \textbf{(b)} the computational domain $\Omega$.  The inner most blue region shows the locations where
the tumor cells will die out within 24 hours.  The outermost orange cells are predicted to die out within 72
hours of exposure.  Finally, the deep red cells towards the outside of the regions do not die after up to
72 hours of exposure.}
\label{fig:frac1}
\end{figure}

In Fig. \ref{fig:frac1}, we present the apoptotic region induced by the initial bump function within the tumor region
$\mathbb{T}$ (Fig. \ref{fig:frac1}a) and the entire computational domain $\Omega$ (Fig. Fig. \ref{fig:frac1}b); that is,
the region of cells expected to die within an exposure time $\tau$ simply due to the initial condition before any
diffusive spreading has occurred.  In the figure, the outermost red region represents the location of the cells where
the drug concentration is below any threshold value.  The innermost dark blue region shows the locations where the
concentration is above the threshold value $u_{T}(24)$, and the heat map illustrates regions where the concentration
of the bump function is above the respective threshold values.

\begin{figure}[htbp]
\centering
\stackinset{l}{0mm}{t}{4mm}{{\color{white}\textbf{\large (a)}}}{\includegraphics[width=0.49\textwidth]{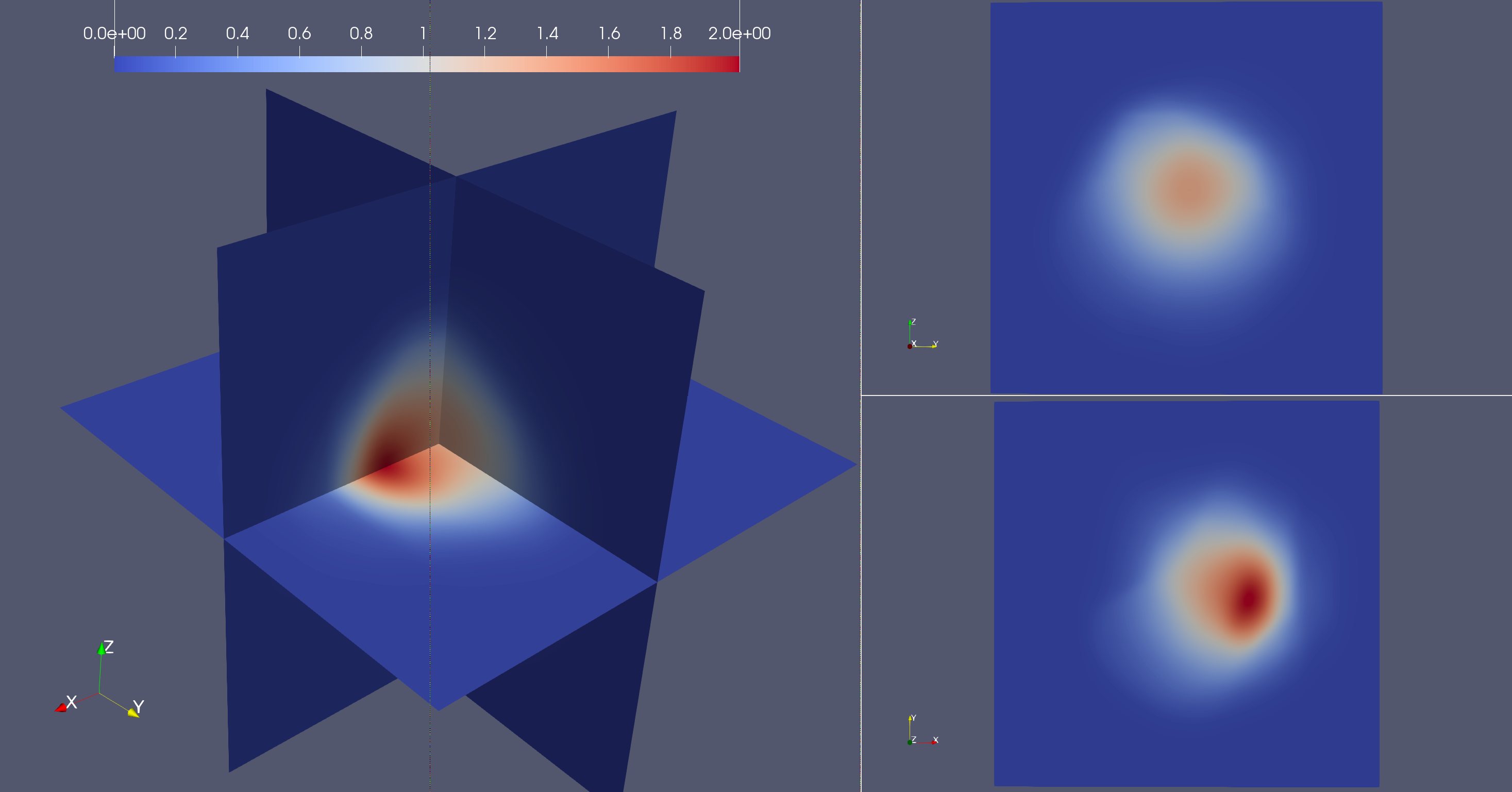}}
\stackinset{l}{0mm}{t}{4mm}{{\color{white}\textbf{\large (b)}}}{\includegraphics[width=0.49\textwidth]{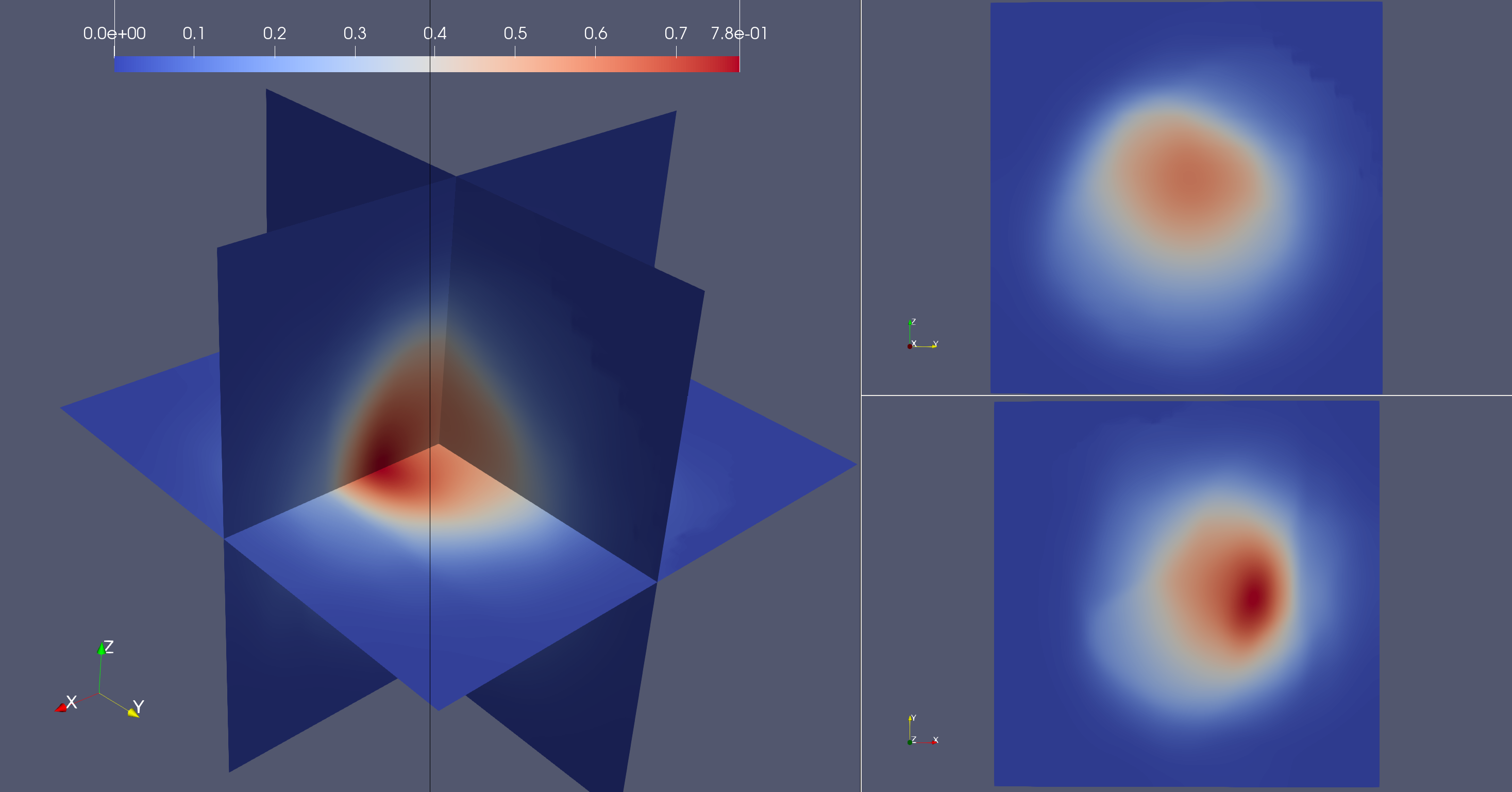}}
\stackinset{l}{0mm}{t}{4mm}{{\color{white}\textbf{\large (c)}}}{\includegraphics[width=0.49\textwidth]{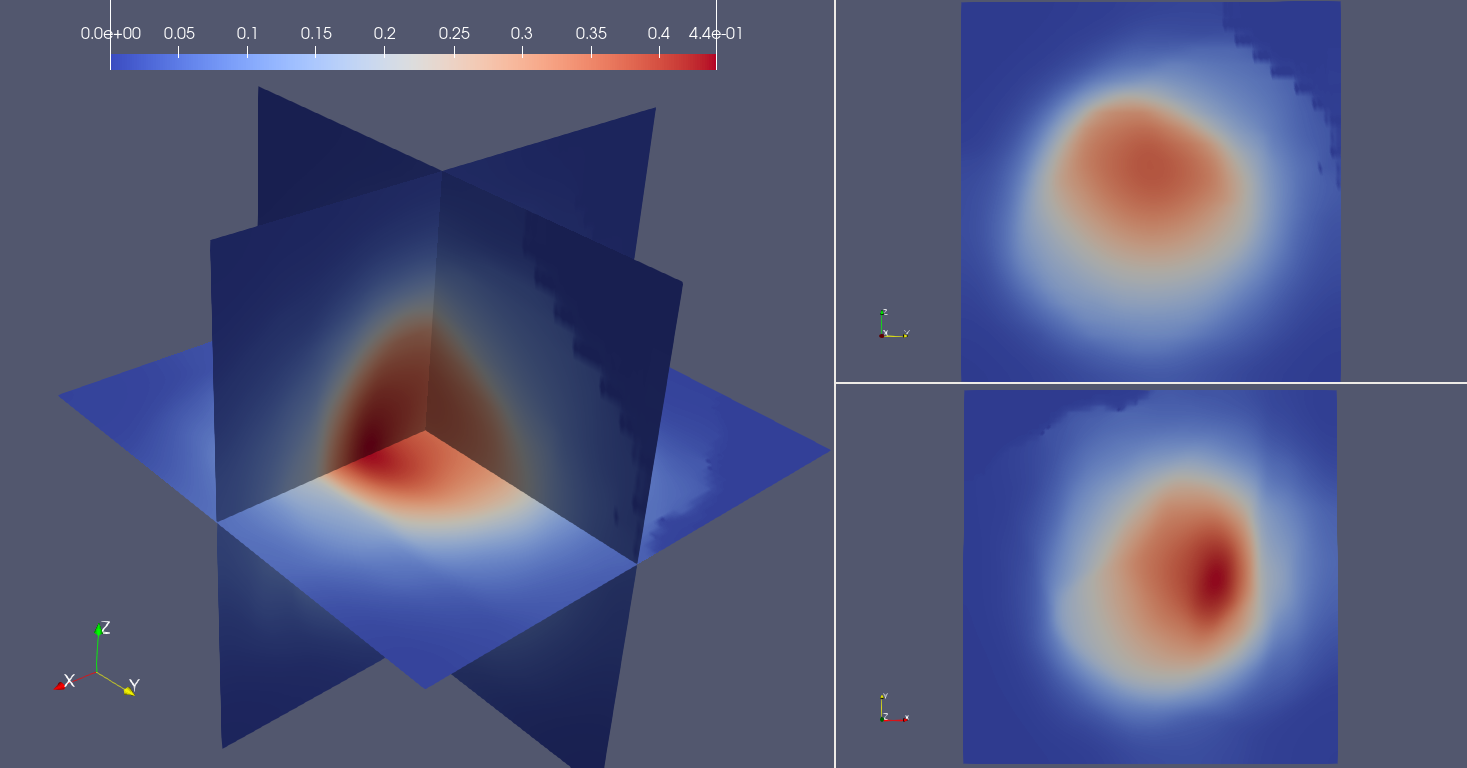}}
\stackinset{l}{0mm}{t}{4mm}{{\color{white}\textbf{\large (d)}}}{\includegraphics[width=0.49\textwidth]{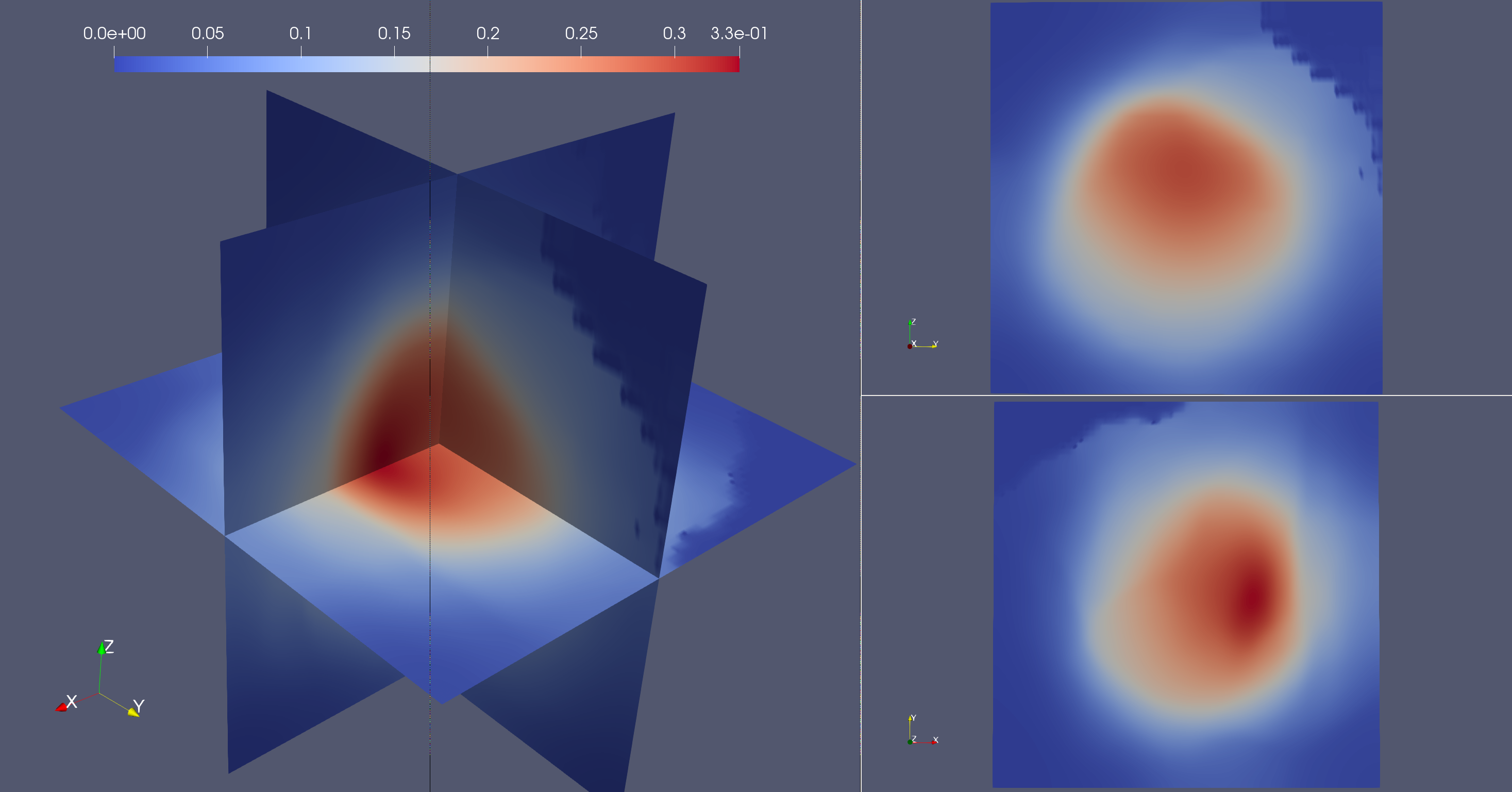}}
\caption{Concentration profiles for an initial injection of $U_{0}=1.5$ $\mu\text{M}$ centered at $\x_c = (0.5,0,0)$
at $t=0.2,0.5,0.8,1$, [\textbf{(a) - \textbf{(d)}}]  respectively.  The heat map is scaled to the visible range for each
figure; i.e., the intensities are with respect to the concentration distribution in each figure independently of the other
figures.  Each figure is displayed with the intersection of three representative slices and two separate views of the frontal,
$yz$-plane, and transverse $xy$-plane.  The inhomogeneous - anisotropic nature of the diffusion can be observed in the
planar slices accompanying the 3-D figures.}
\label{fig:cprofs}
\end{figure}

As the drug diffuses across the computational domain, $\Omega$, more cells are exposed to the drug
over the course of the simulation.  Figure \ref{fig:cprofs} shows the planar projections of the concentration
profiles as a heat map at increasing simulation times.  Red represents a higher concentration of drugs and
deep blue represents a concentration of zero.  To improve the visibility of simulations, we scale the data to
visible data range.  The inhomogeneous - anisotropic nature of the diffusion can be clearly observed from
the concentration profiles in Fig. \ref{fig:cprofs}.


Now, we can compute the concentration levels at particular simulations times and compare them to the threshold
values.  For a particular simulation time, $t_n$, the volume fraction of the tumor where the concentration is
above the threshold, $u_T(\tau)$, for the first time is calculated and summed with that of all previous times
$t_i$ where $i < n$, using \eqref{Eq: Beth}.  When this is done for the final time, $t_n = T$, which is chosen
to be large enough that all of the drug mixture leaks away after this time, then we have our apoptosis fraction,
$\aleph(\tau)$ in \eqref{Eq: Aleph}.  We report the exact fractions, $\beth(\tau,t_n)$, from \eqref{Eq: Beth} in
Table \ref{tab:aptable}.  Note that the fourth column is the final values that indicate the percentage of the tumor
cells killed, $\aleph(\tau)$.

\begin{table}[htbp]
\centering
\caption{Convergence of apoptosis fraction approximation for $U_{0}=1.5$ $\mu\text{M}$}
\begin{tabular}{|c|c|c|c|c|c|}
 \hline
  & $t_n=0.2$ & $t_n=0.5$ & $t_n=0.8$ & $t_n=1$ & $t_n=1.2$  \\
 \hline
 $\beth(24,t_n)$ & 0.498408 & 0.528229  & 0.528229 & 0.528229 & 0.528229\\
 \hline
 $\beth(48,t_n)$ & 0.698709 & 0.814072   & 0.864979  & 0.873871 &0.877210\\
 \hline
 $\beth(72,t_n)$ & 0.750721 & 0.878613 & 0.909647 & 0.915700 &0.921299\\
 \hline
\end{tabular}
\label{tab:aptable}
\end{table}

\begin{figure}[htbp]
\centering
\stackinset{l}{0mm}{t}{4mm}{{\color{white}\textbf{\large (a)}}}{\includegraphics[width=.5\textwidth]{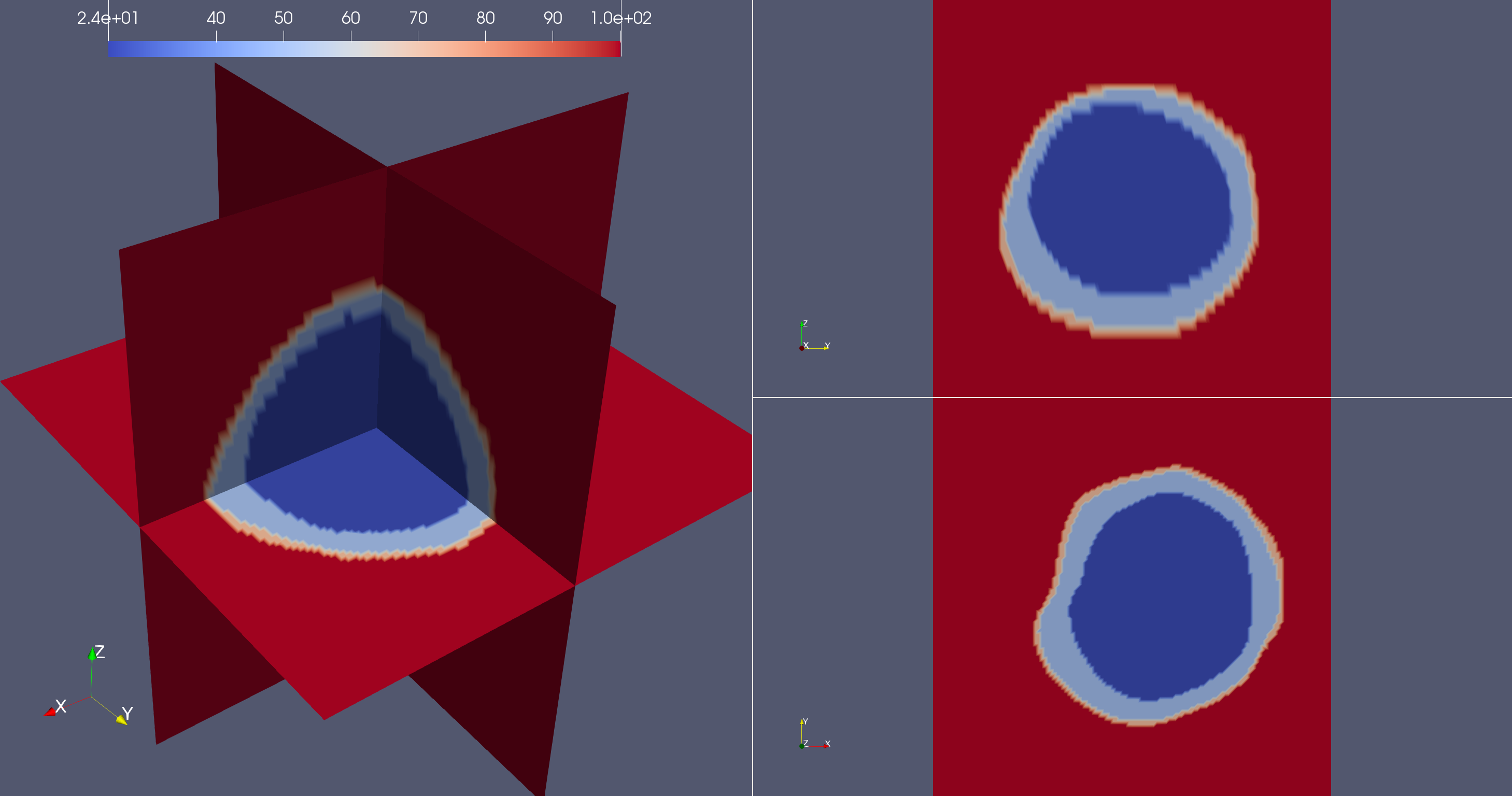}}
\stackinset{l}{0mm}{t}{4mm}{{\color{white}\textbf{\large (b)}}}{\includegraphics[width=.485\textwidth]{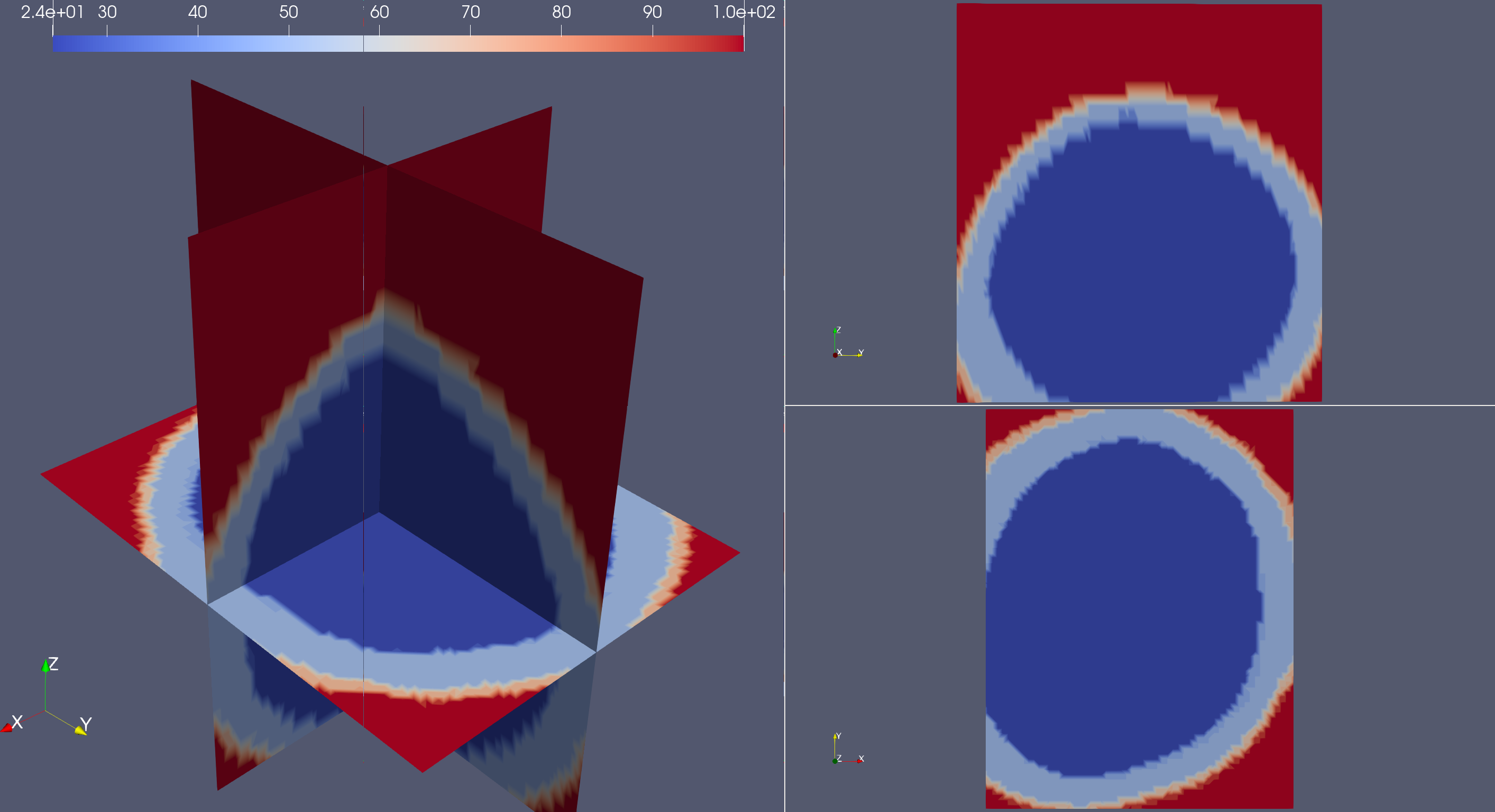}}
\stackinset{l}{0mm}{t}{4mm}{{\color{white}\textbf{\large (c)}}}{\includegraphics[width=.5\textwidth]{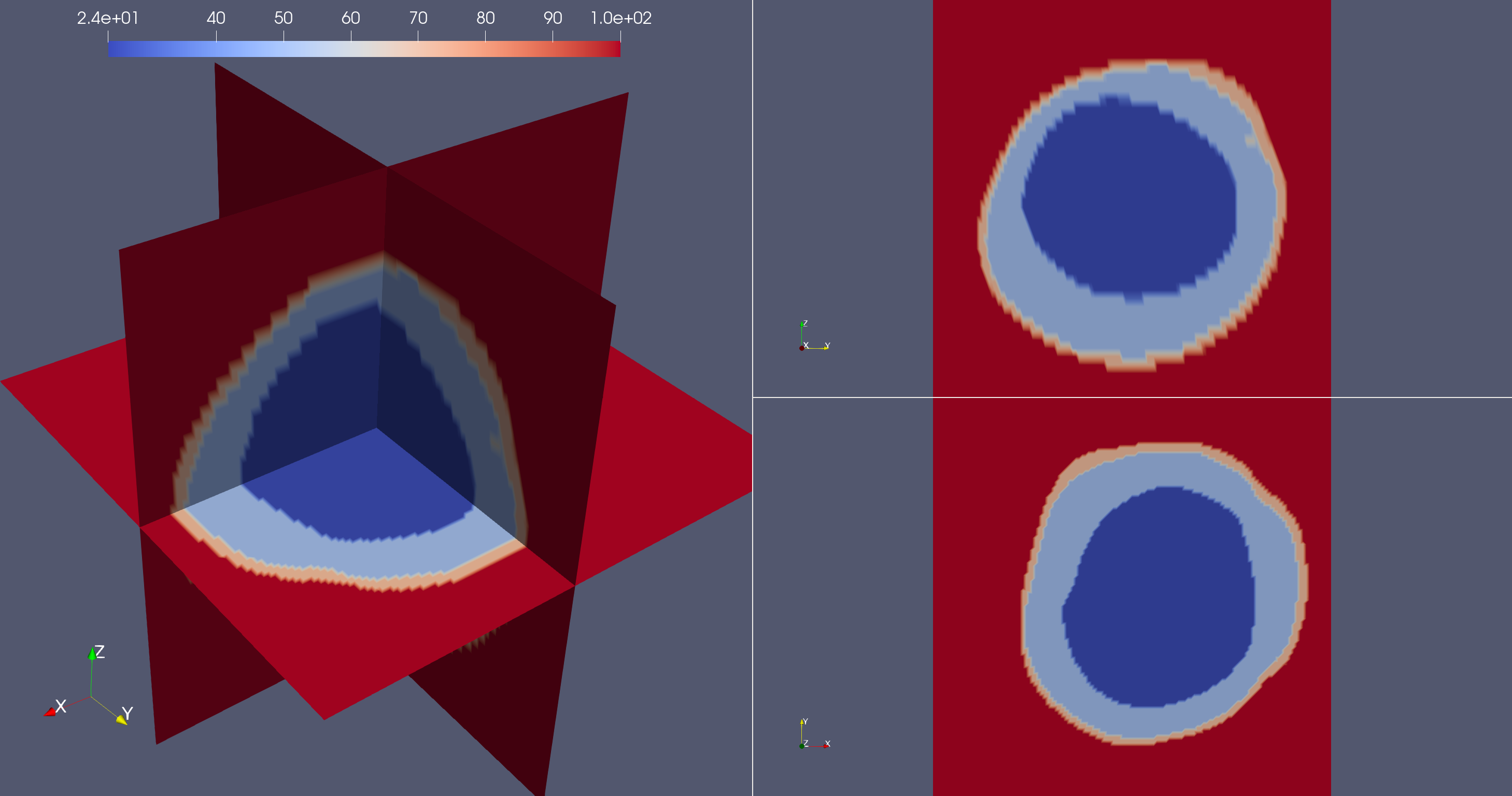}}
\stackinset{l}{0mm}{t}{4mm}{{\color{white}\textbf{\large (d)}}}{\includegraphics[width=.485\textwidth]{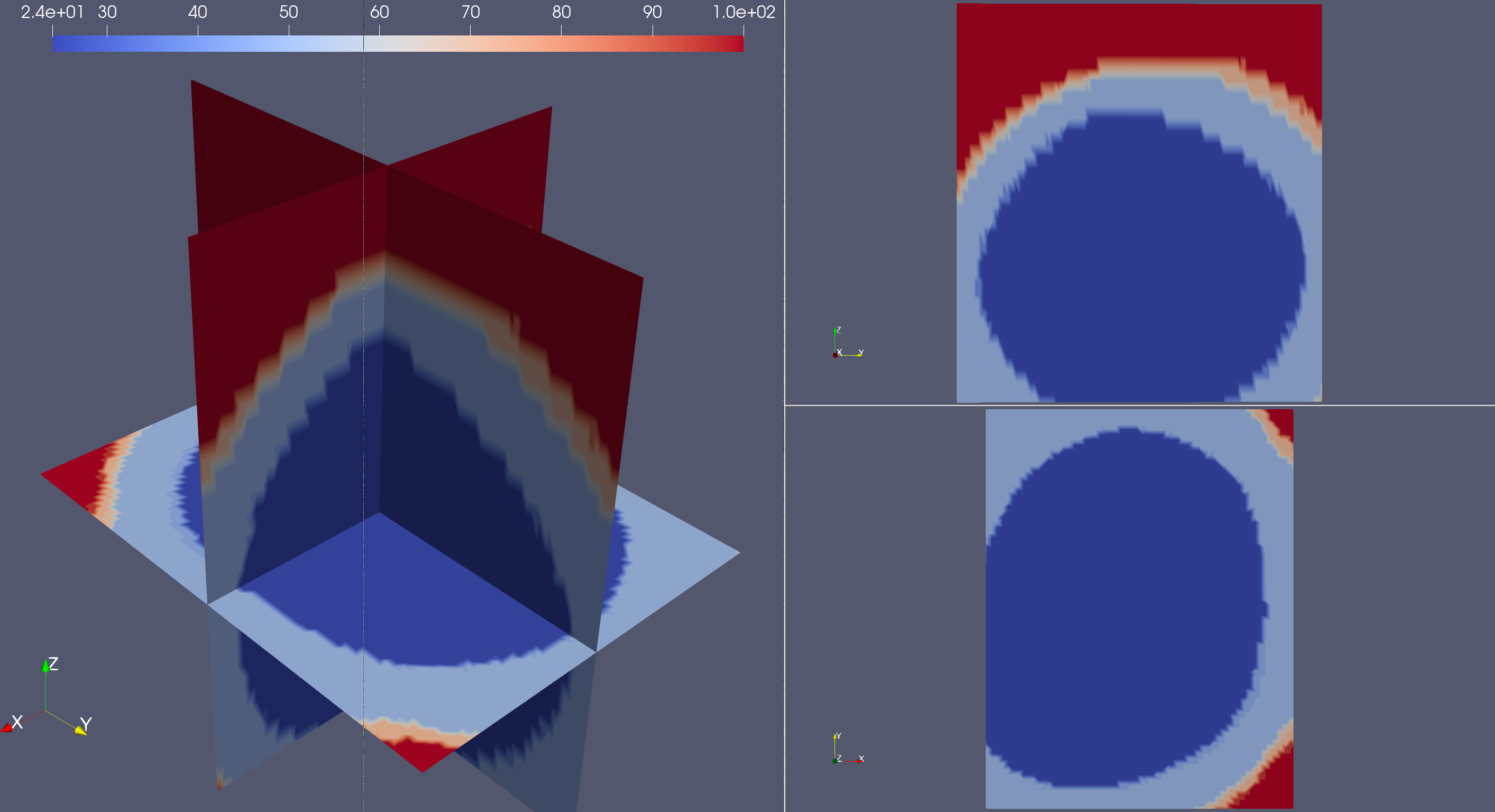}}
\stackinset{l}{0mm}{t}{4mm}{{\color{white}\textbf{\large (e)}}}{\includegraphics[width=.5\textwidth]{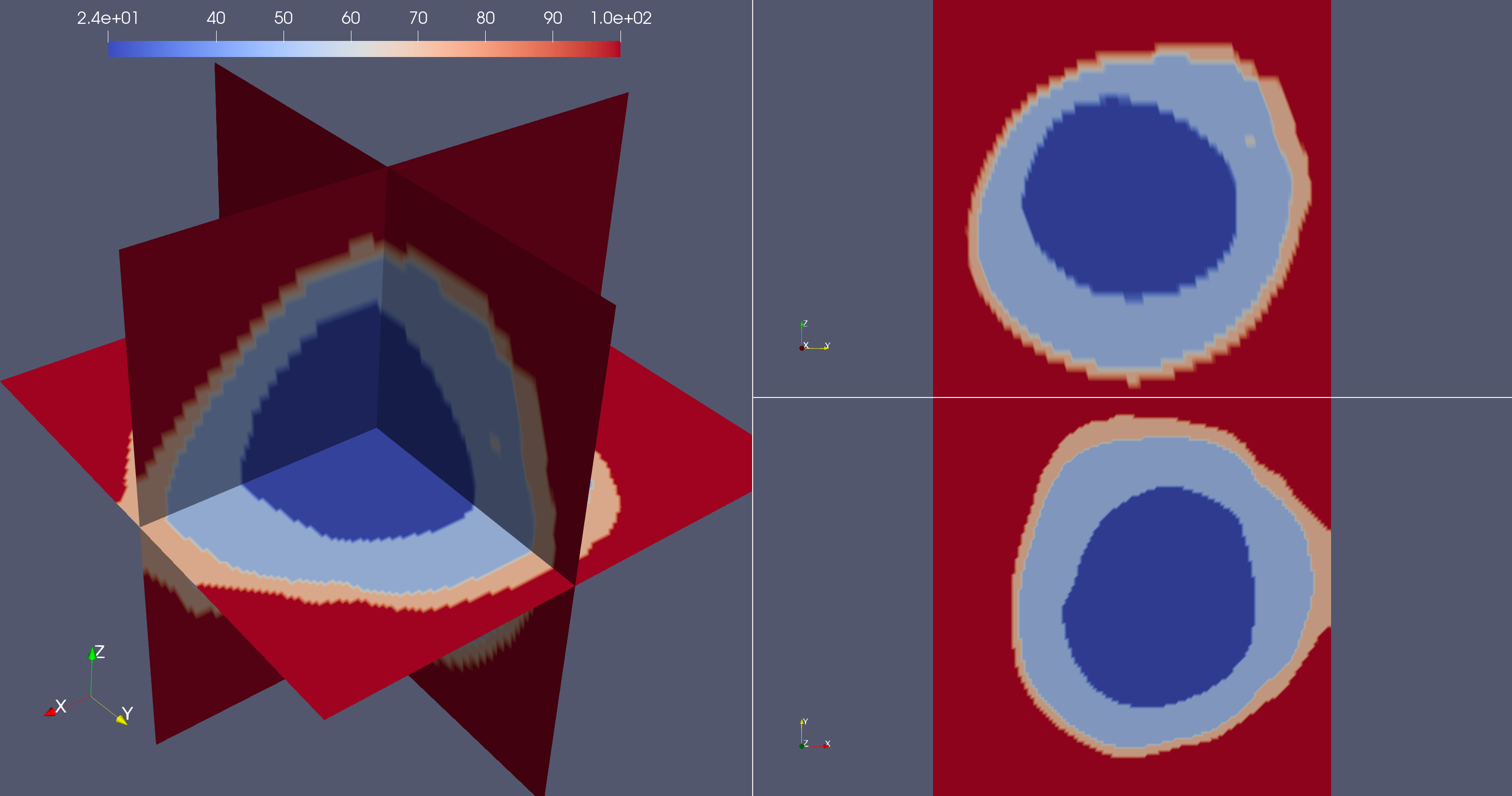}}
\stackinset{l}{0mm}{t}{4mm}{{\color{white}\textbf{\large (f)}}}{\includegraphics[width=.485\textwidth]{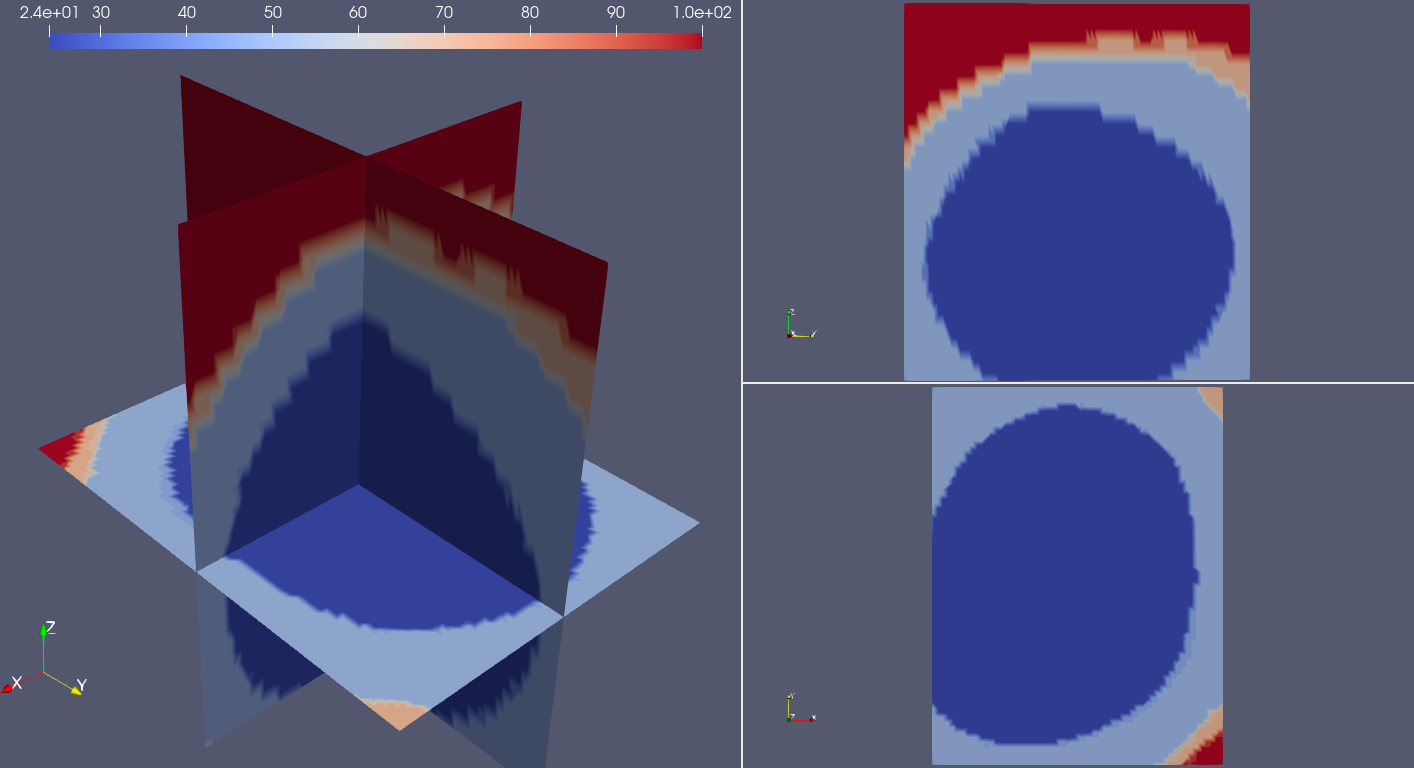}}
\stackinset{l}{0mm}{t}{4mm}{{\color{white}\textbf{\large (g)}}}{\includegraphics[width=.5\textwidth]{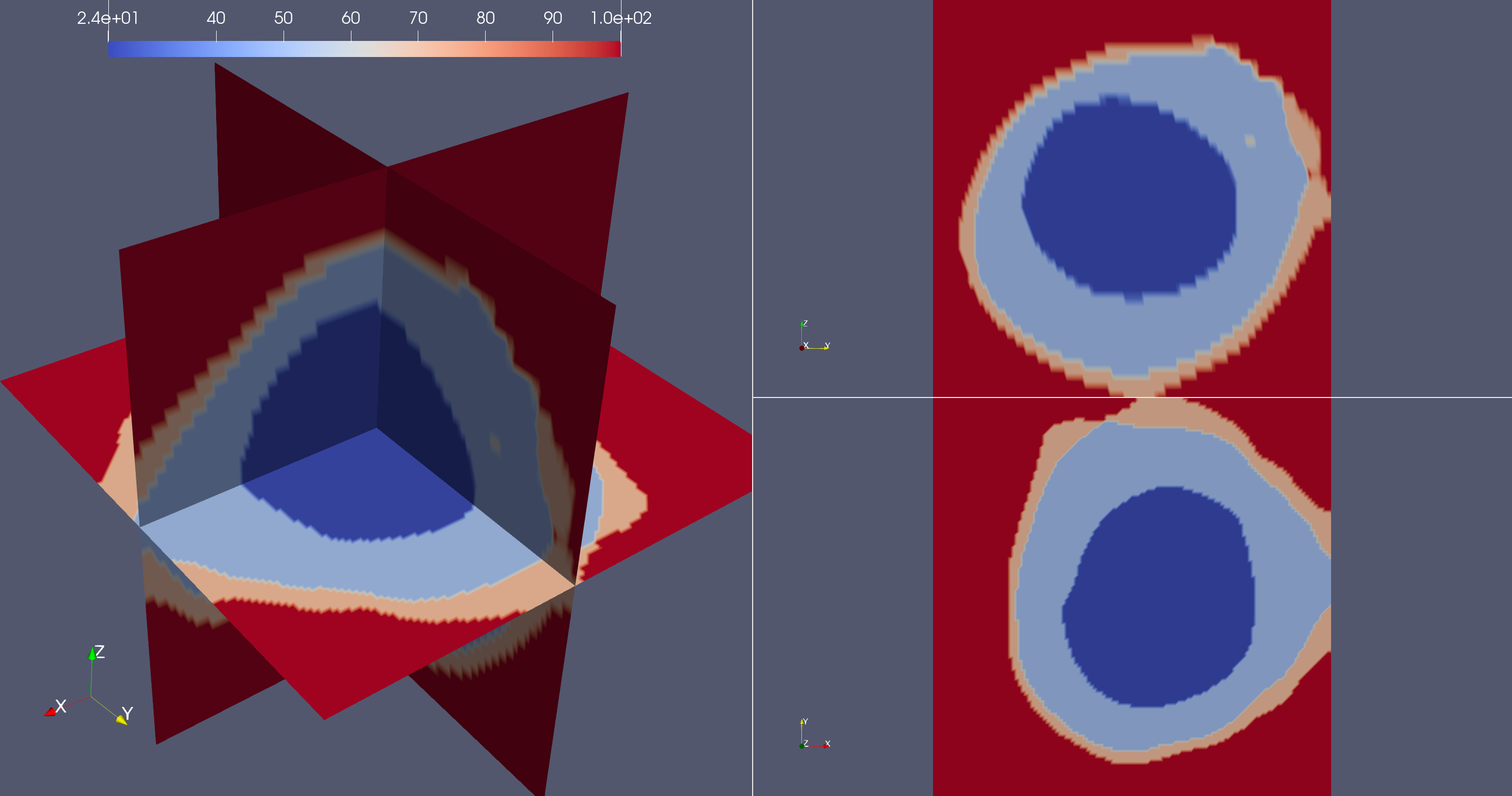}}
\stackinset{l}{0mm}{t}{4mm}{{\color{white}\textbf{\large (h)}}}{\includegraphics[width=.485\textwidth]{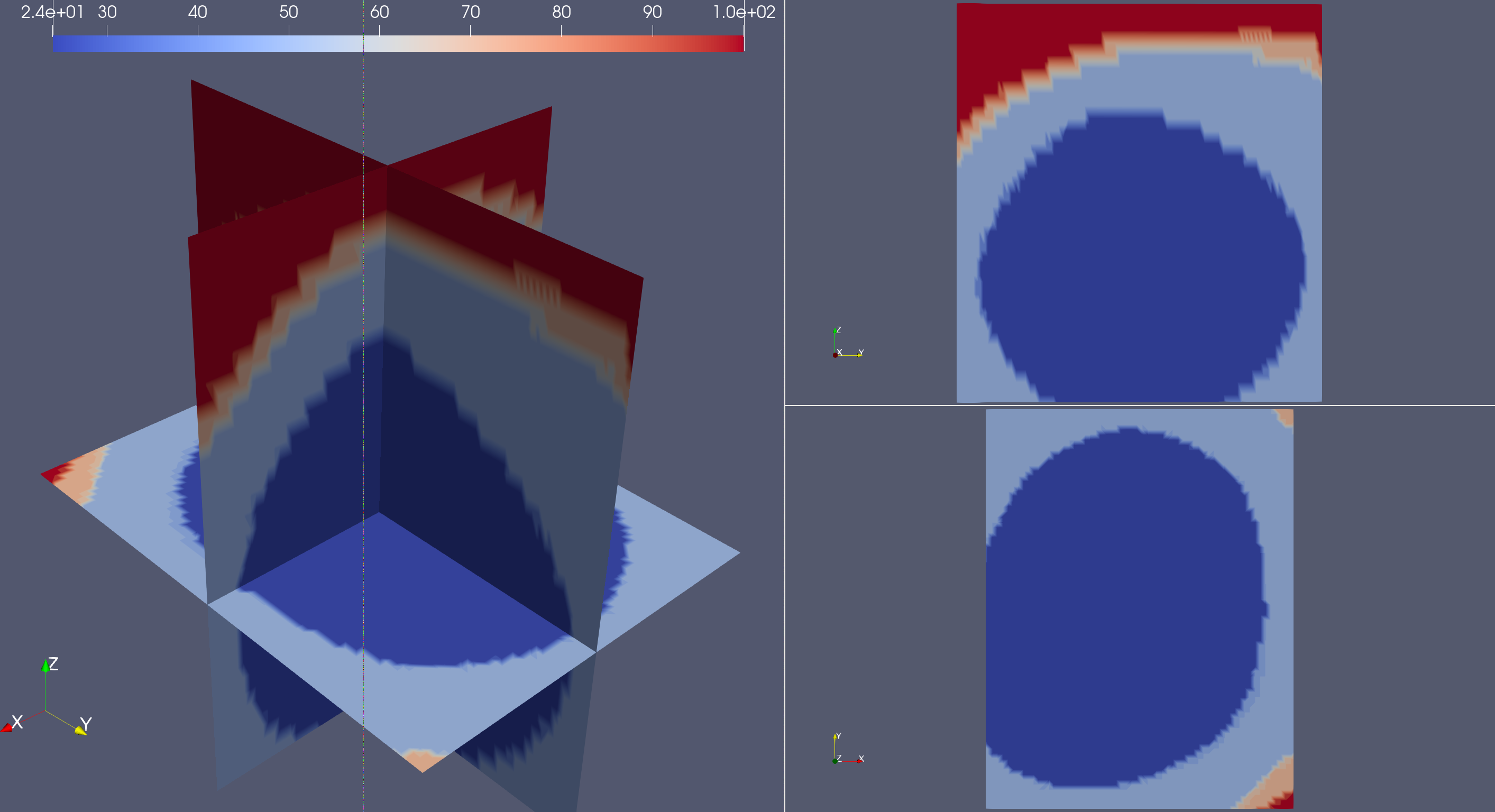}}
\caption{Cells with a drug concentration higher than the given threshold, $\beth(\tau,t_n)$, at simulation times
$t_n = 0.2,\, 0.5,\, 0.8,\, 1$ after an initial injection of $U_{0}=1.5$ $\mu\text{M}$ centered at $\x_c = (0.5,0,0)$.
The column on the left, \textbf{(a)}, \textbf{(c)}, \textbf{(e)}, \textbf{(g)}, shows the entire computational domain,
$\Omega$, and the column on the right, \textbf{(b)}, \textbf{(d)}, \textbf{(f)}, \textbf{(h)}, shows the tumor
region, $\mathbb{T}$.  Each figure is displayed with the intersection of three representative slices and two separate views
of the frontal, $yz$-plane, and transverse $xy$-plane.  It is observed that the number of cells exposed to a sufficient
amount of drugs is higher on transverse, compared to the sagital, $xz$-plane, and the frontal plane.  The inner most blue
region shows the locations where $u(\x,t) > u_T(\tau = 24 \text{hours})$ for any $t = t_n$.  The outermost orange
cells correspond to $\tau = 72$ hours.  Finally, the deep red cells towards the outside of the regions has concentrations
$u(\x,t) < u_T(\tau)$ for all $t$ and $\tau$.}
\label{fig:dprofs}
\end{figure}

\begin{figure}[htbp]
\centering
\stackinset{r}{0mm}{t}{2mm}{{\color{white}\textbf{\large (a)}}}{\includegraphics[width=0.49\textwidth]{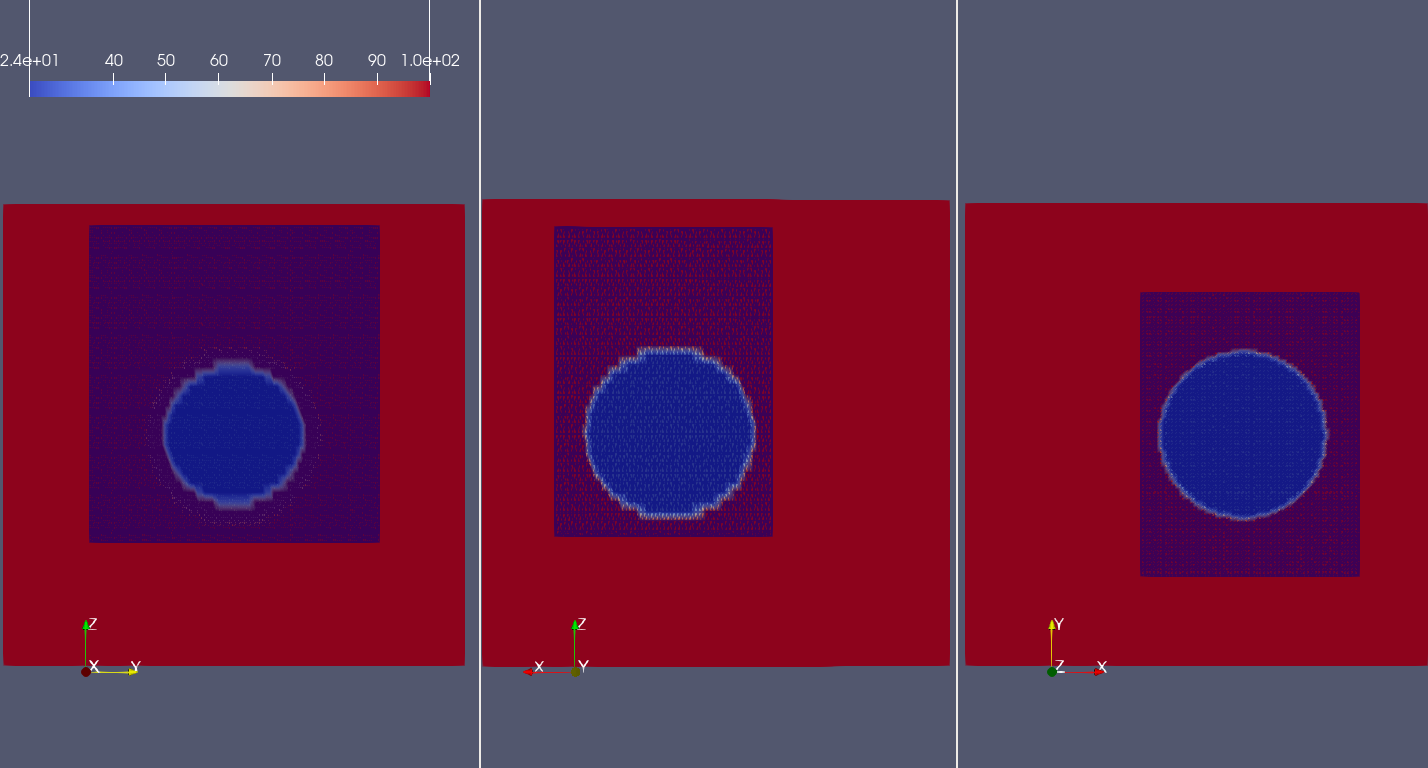}}
\stackinset{r}{0mm}{t}{2mm}{{\color{white}\textbf{\large (b)}}}{\includegraphics[width=0.49\textwidth]{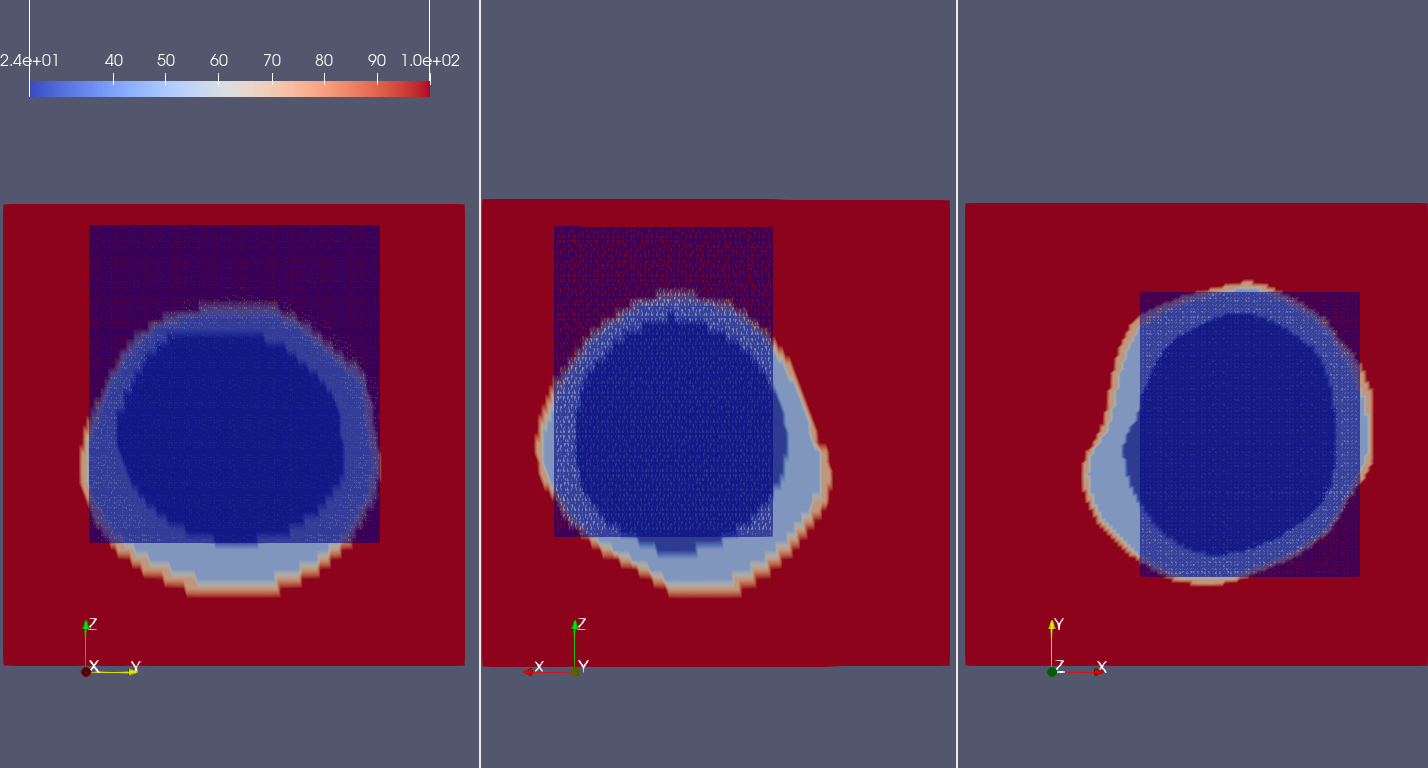}}
\stackinset{r}{0mm}{t}{2mm}{{\color{white}\textbf{\large (c)}}}{\includegraphics[width=0.49\textwidth]{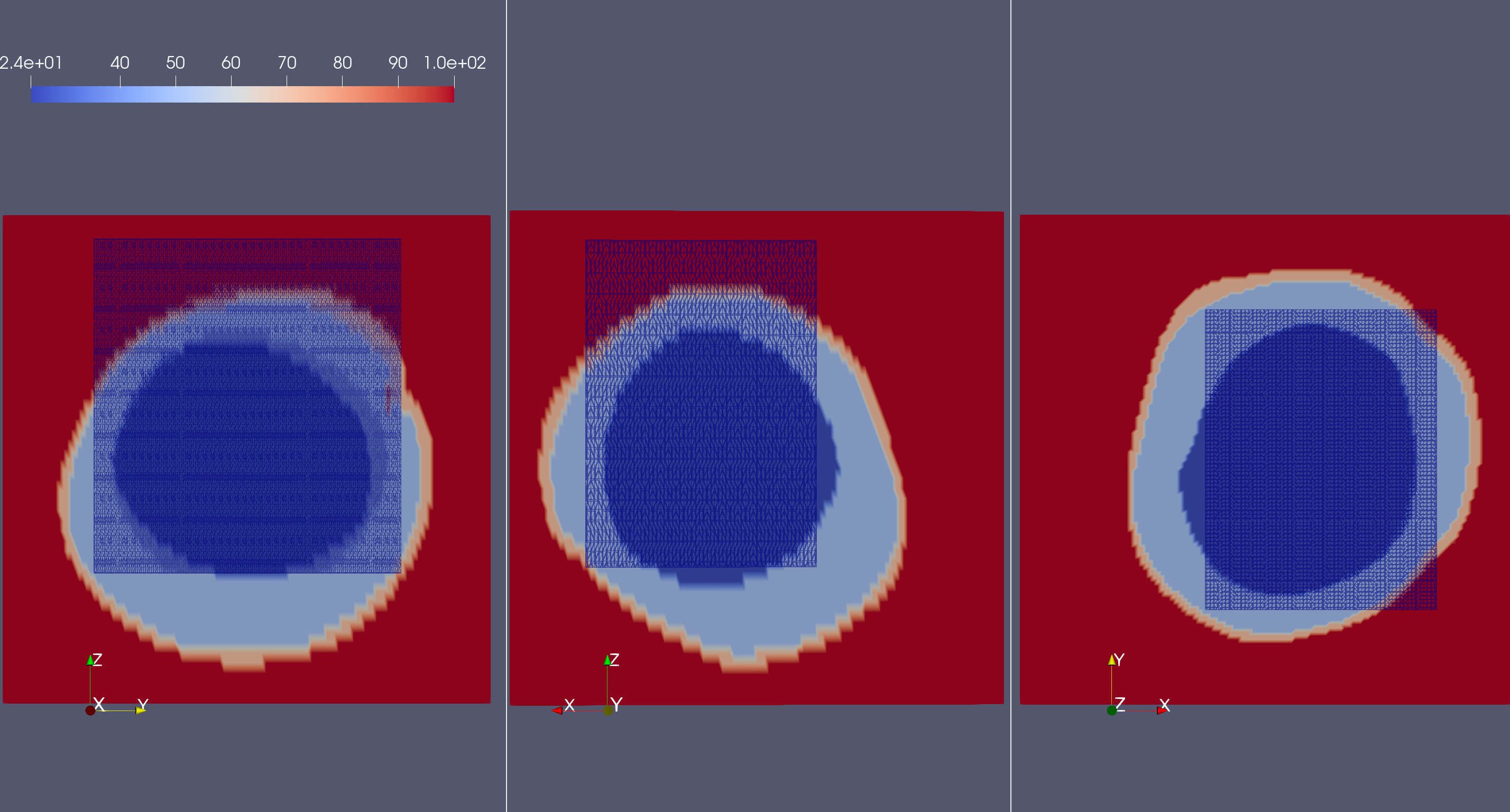}}
\stackinset{r}{0mm}{t}{2mm}{{\color{white}\textbf{\large (d)}}}{\includegraphics[width=0.49\textwidth]{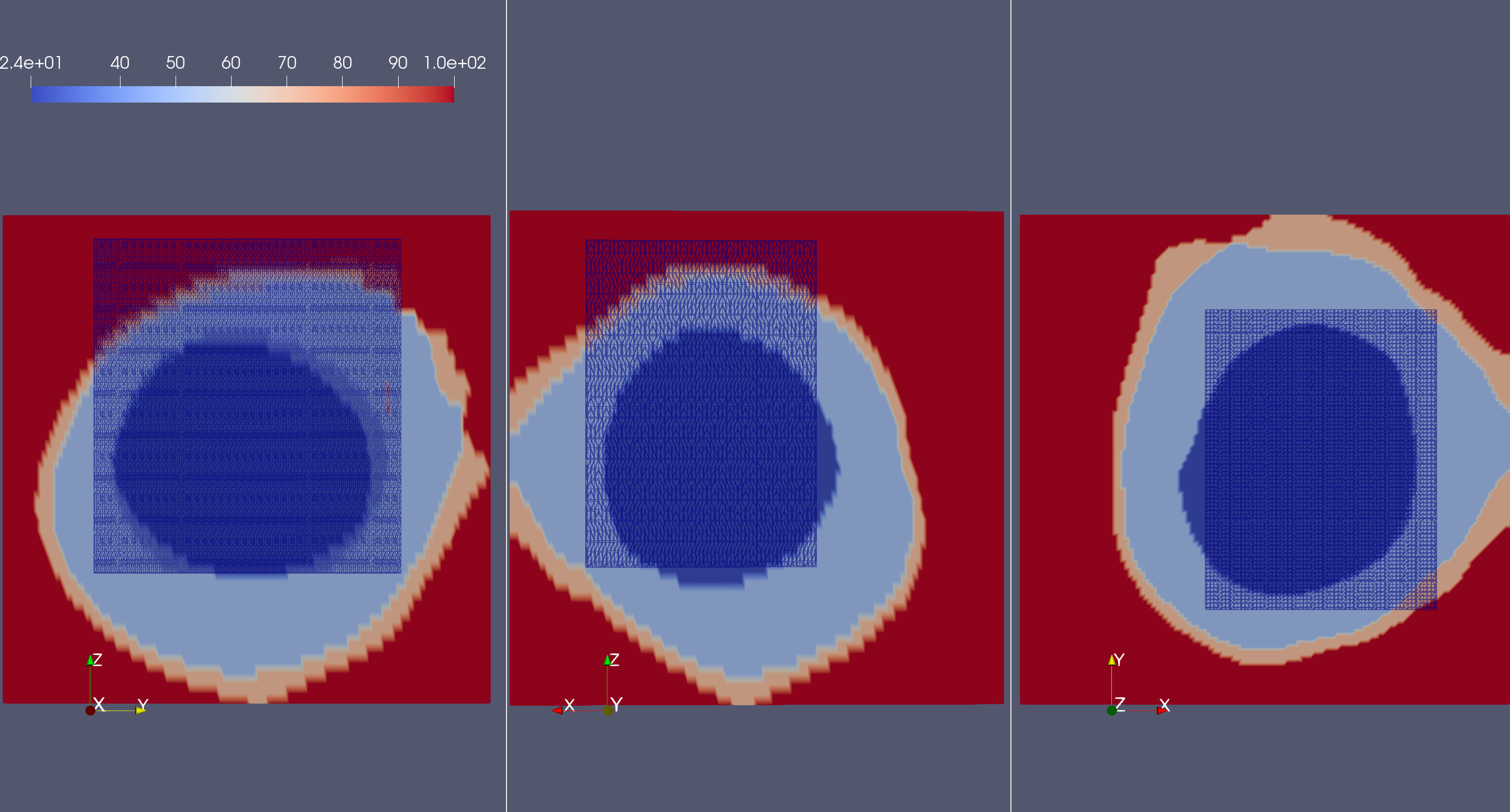}}
\caption{A visual demonstration of toxicity.  The effect of the threshold is extended beyond the tumor region,
$\mathbb{T}$, (the dark blue rectangular mesh), and the heat map is calculated in the $yz$, $xz$, and $xy$-planes 
from \eqref{Eq: Beth} at simulation times $t_n = 0,\, 0.2,\, 0.5,\, 1$, [\textbf{(a) - \textbf{(d)}}] respectively.  It
is observed that most of the drug mixture eventually diffuses outside of the tumor region. Thus, significant portions of
healthy cells may potentially be subjected to toxicity.  The inner most blue region shows the locations where
$u(\x,t) > u_T(\tau = 24 \text{hours})$ for any $t = t_n$.  The outermost orange cells correspond to $\tau = 72$ hours.
Finally, the deep red cells towards the outside of the regions has concentrations $u(\x,t) < u_T(\tau)$ for all $t$
and $\tau$.}
\label{fig:toxi}
\end{figure}

The effects of the inhomogeneous - anisotropic diffusion on apoptosis can be observed in Fig. \ref{fig:dprofs}.
For example, the drug efficacy is significantly higher on the transverse plane compared to the sagital and frontal
planes.  This can be seen in Fig. \ref{fig:dprofs}h where the apoptotic region covers almost the entire transverse plane,
while nontrivial portions of the frontal plane remains unscathed.  Moreover, even though relatively high apoptosis
fractions are achieved in the tumor region, $\mathbb{T}$, it is observed in Fig. \ref{fig:toxi} that significant amounts
of the drug leak out of $\mathbb{T}$. So our results indicate that considerable portions of the healthy cells are subject
to toxicity.  We can visually inspect the leakage of the drug by superimposing the representative slices of the
computational domain $\Omega$ and the sub-region $\mathbb{T}$ (Fig. \ref{fig:toxi}).

\begin{figure}[htbp]
\centering  
\includegraphics[width=0.9\textwidth]{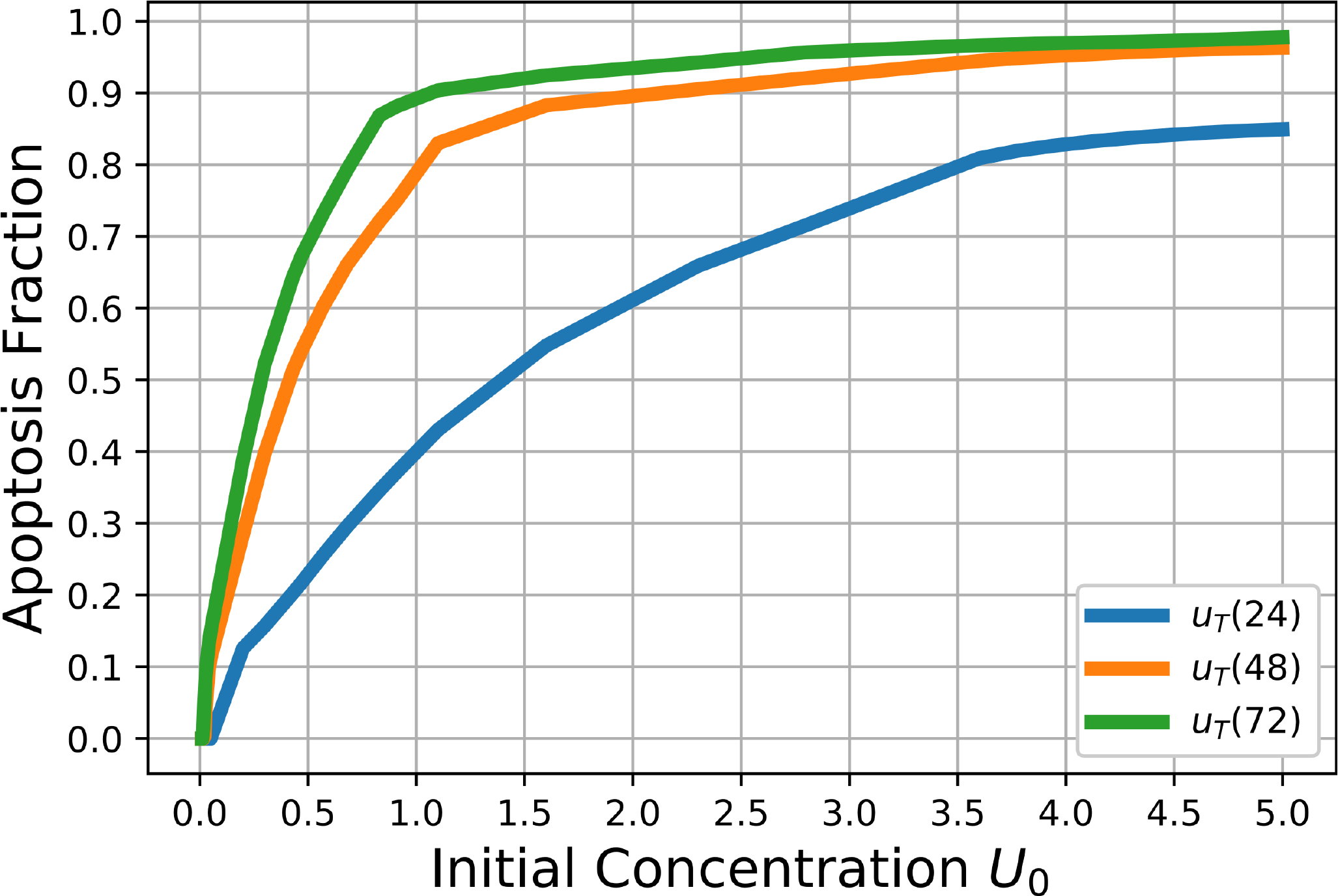}
\caption{Dose-response curves (24, 48, and 72 hours from bottom to top) produced from 31 different initial injections
varying from over the interval $[0.01,5]$.  It can be seen that the apoptosis fraction starts to asymptote beyond
$U_{0}=2$ for 48 and 72 hours exposure options.}
\label{fig:dose_res}
\end{figure}

It is observed in Table \ref{tab:aptable} that the current configuration of the problem leads to partial ablation in
the tumor region since 92 \% of the region dies after 72 hours.  In fact, as shown in the dose-response curves in
Fig. \ref{fig:dose_res}, the apoptosis fraction seemingly asymptotes as we increase the amount of initial injection
because the interaction of the drug with several obstacles, local inhomogeneities, and leakage, do not allow 
sufficient concentrations to diffuse to every part of the tumor region. Thus, we can conclude that our mechanistic
model is capable of reflecting the diffusion patterns mostly governed by location-wise structural differences.

Considering the internal injection scenario, several ways can be proposed to improve the efficacy of the drug without
changing its fluidic properties.  An obvious way may be to use a higher initial concentration, $U_0$.  In fact, we can
create dose-response curves for a broader view of the effects of various initial concentrations $U_{0}$ on the final
apoptosis fractions.  Figure \ref{fig:dose_res} demonstrates the final percentages of the tumor that is killed for 31
different initial injections varying from $U_0 = 0.01$ $\mu\text{M}$ to $U_0 = 5$ $\mu\text{M}$. Dose-response
curves reveal that even though the initial injection is excessively elevated, it is not possible to reach a complete
tumor ablation. Indeed it is observed that apoptosis fractions barely improve after approximately $U_{0}=2$ $\mu\text{M}$
for $\tau = 48$ and $72$ hours exposure times. We reported in Fig. \ref{fig:toxi} that even $U_{0}=1.5$ $\mu\text{M}$
causes the presence of significant concentrations of drugs outside of the tumor region $\mathbb{T}$.  Thus we
can conclude that the use of initial drug concentrations outside of a certain range is prohibitive and can severely
contribute to toxicity. 

\begin{figure}[htbp]
\centering
\stackinset{l}{5.5mm}{t}{1mm}{{\textbf{\large (a)}}}{\includegraphics[width=0.32\textwidth]{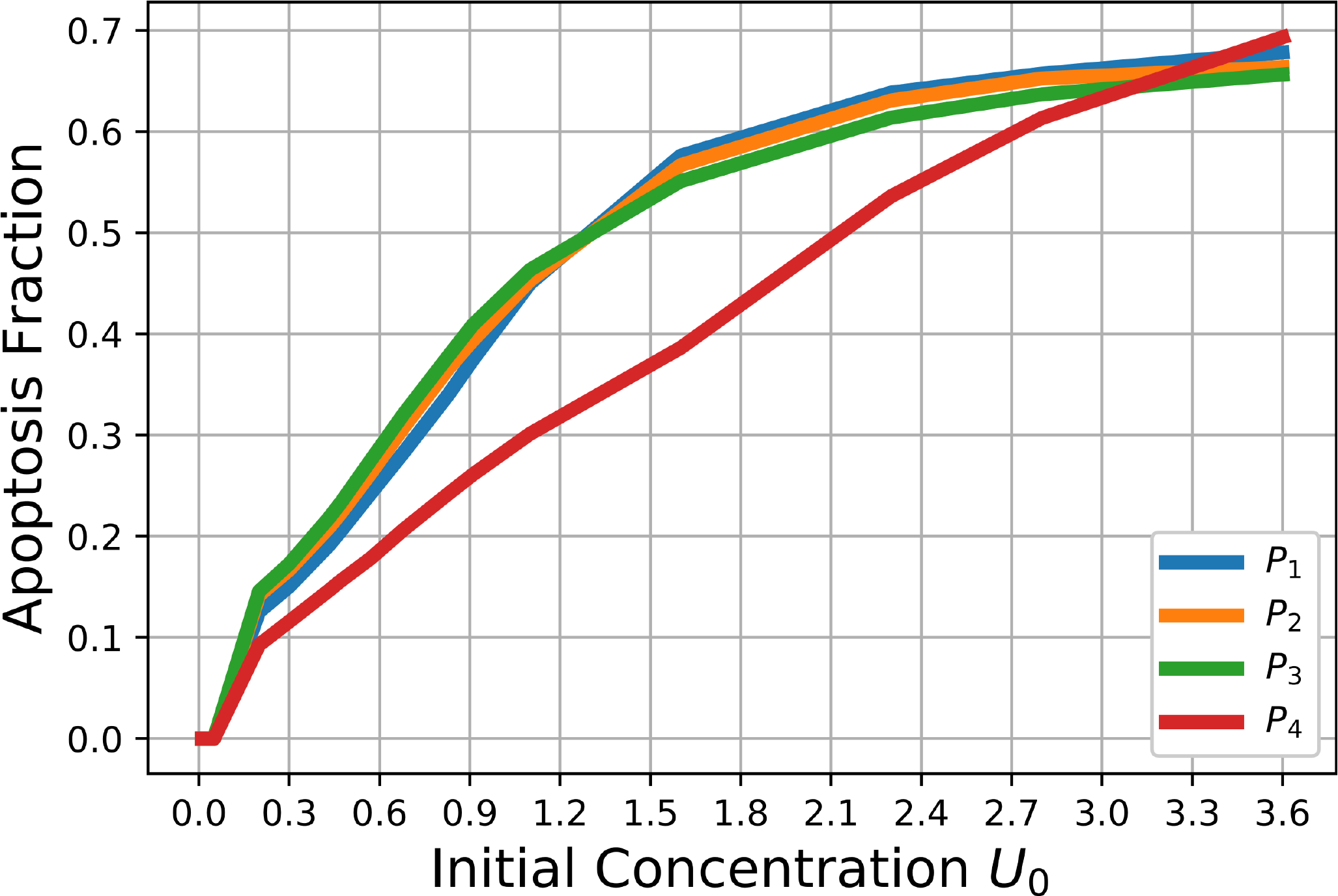}}
\stackinset{l}{5.5mm}{t}{1mm}{{\textbf{\large (b)}}}{\includegraphics[width=0.32\textwidth]{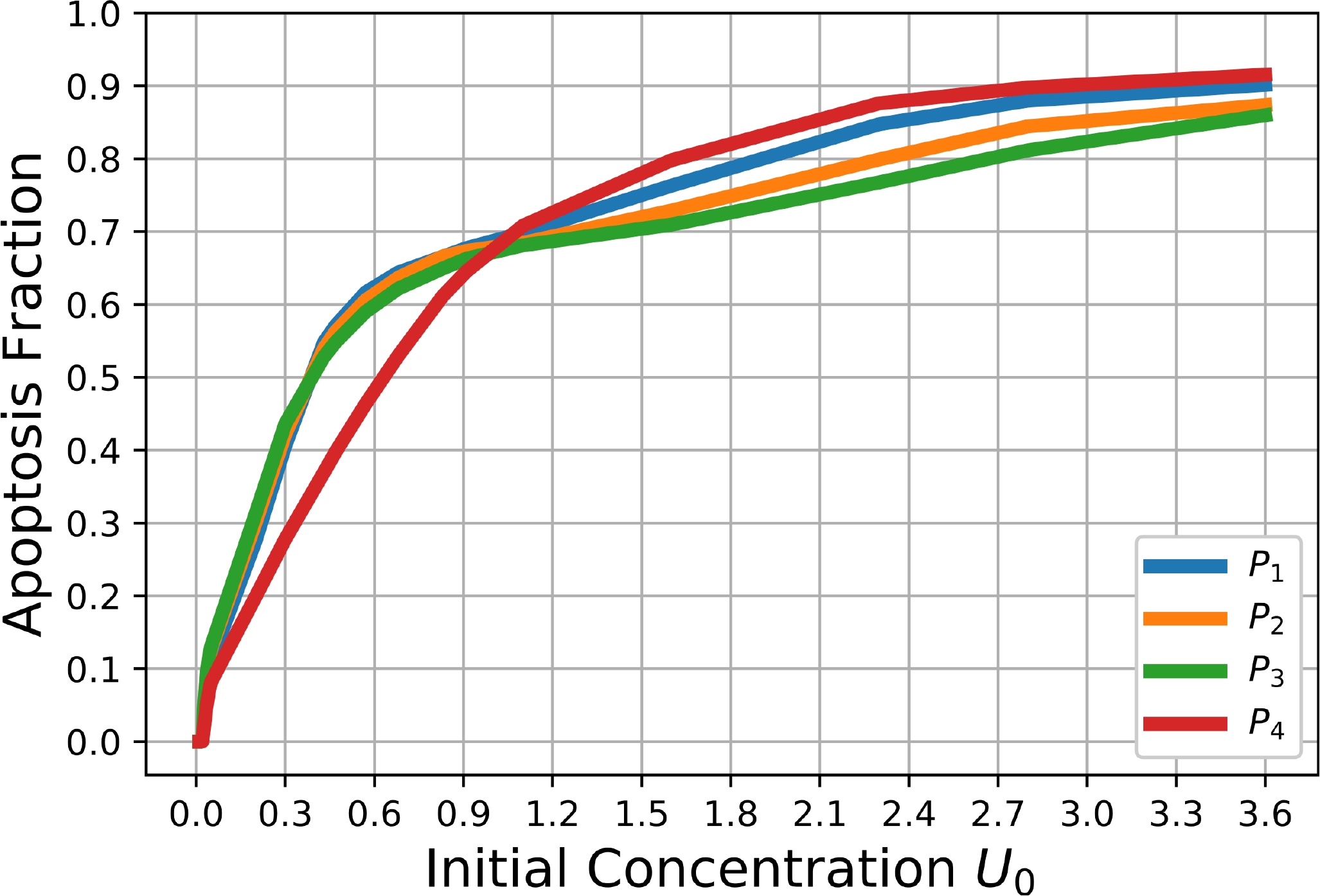}}
\stackinset{l}{5.5mm}{t}{1mm}{{\textbf{\large (c)}}}{\includegraphics[width=0.32\textwidth]{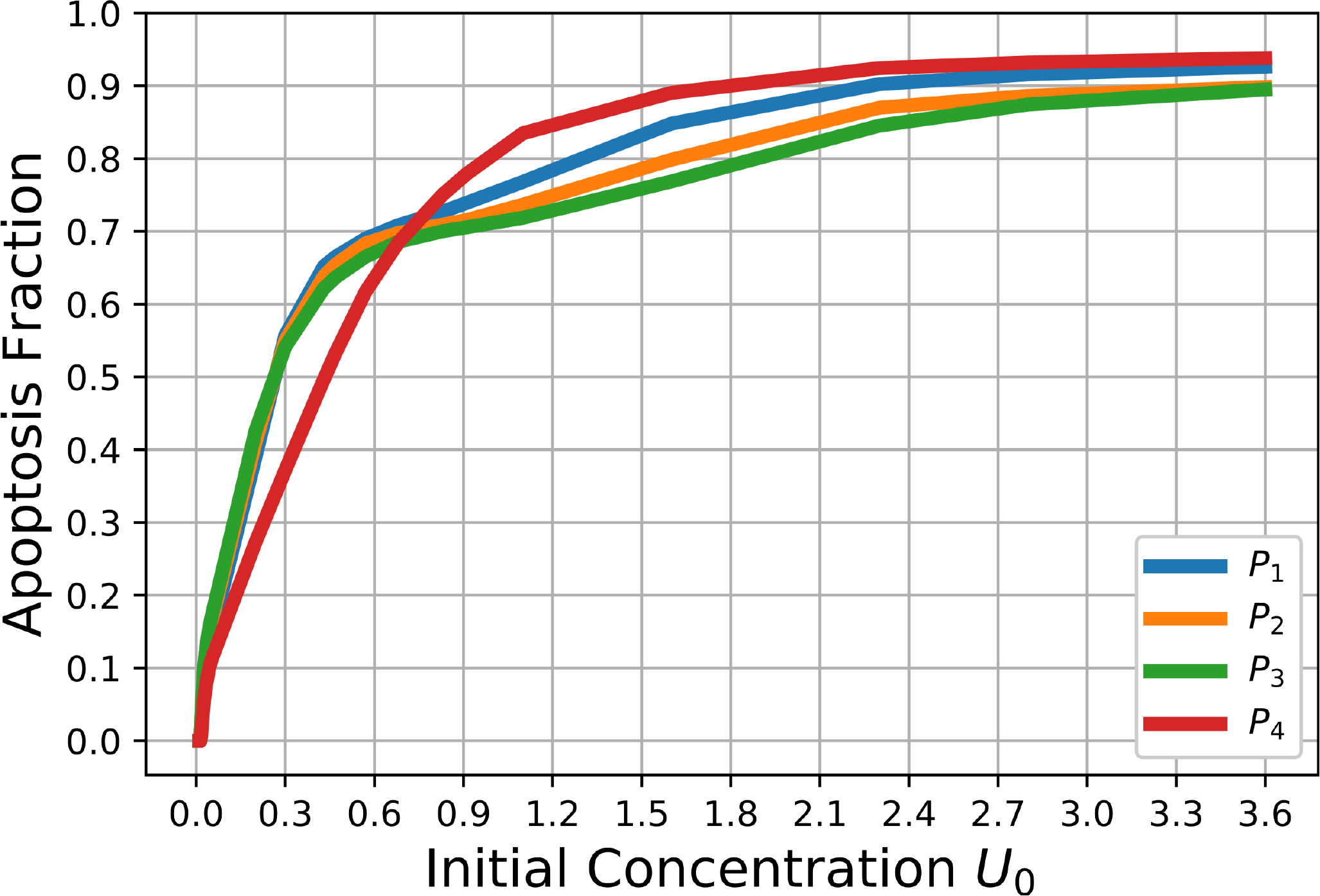}}
\caption{Dose-response curves with 4 different injection points: $P_{1}(0.5,-0.3,-0.5)$, $P_{2}(-0.3,0,-0.6)$,
$P_{3}(-0.3,-0.3,-0.6)$, and $P_{4}(0,1,0)$.  \textbf{(a)}  24 hour dose-response curves.  \textbf{(b)} 
48 hour dose-response curves.  \textbf{(c)}  72 hour dose-response curves.  Although $P_{4}$ is picked as
an intuitively ``poor'' location in terms of distance to the center of the tumor bulk, it yields better efficacy
for 48 and 72 hour exposure times after $U_{0}=1$ $\mu\text{M}$ and $U_{0}=0.7$ $\mu\text{M}$, respectively.}
 \label{fig:4CenterRes}
\end{figure}

Another important parameter effecting the apotosis fraction is the location of injection. Our initial choice as the
center of the bump function is $(0.5,0,0)$, which we pick by visual inspection to be near the center of the tumor.
We found that changing the injection location greatly effects the diffusion pattern, and hence the final apoptosis
fraction. In this sense, an important prediction in this investigation is that a seemingly poor location in $\mathbb{T}$
in terms of the distance from the center of the tumor bulk can yield higher apoptosis fractions than some locations
that are close to the tumor center. To illustrate this phenomenon,  in Fig. \ref{fig:4CenterRes}, we choose four
different injections points $P_{1}(0.5,-0.3,-0.5)$, $P_{2}(-0.3,0,-0.6)$, $P_{3}(-0.3,-0.3,-0.6)$, and $P_{4}(0,1,0)$,
and create dose response curves for each location.  Our results demonstrate that finding the optimal injection location
to maximize apoptosis fractions is quite an unpredictable process and cannot be achieved with a basic visual inspection.
For example, we pick $P_{4}$ to be the ``poor'' location considering its position relative to the center of the tumor,
and expected to get lower apoptosis fractions compared to the other locations, but Fig. \ref{fig:4CenterRes} shows a
different outcome.  Although   $P_{1}$, $P_{2}$, and $P_{3}$ yield relatively similar patterns, $P_{4}$ produces a
better results for 48 and 72 exposure times after $U_{0}=0.8$. However, efficacy is remarkably low for $P_1$ comparing
to others if one opts to measure the apoptosis after 24 hours. We should also note that similar to the initial findings,
none of these configurations lead to full ablation in the tumor region. 

\begin{figure}[htbp]
\centering  
%
\stackinset{l}{0mm}{t}{1mm}{{\color{white}\textbf{\large (a)}}}{\includegraphics[width = 0.9\textwidth]{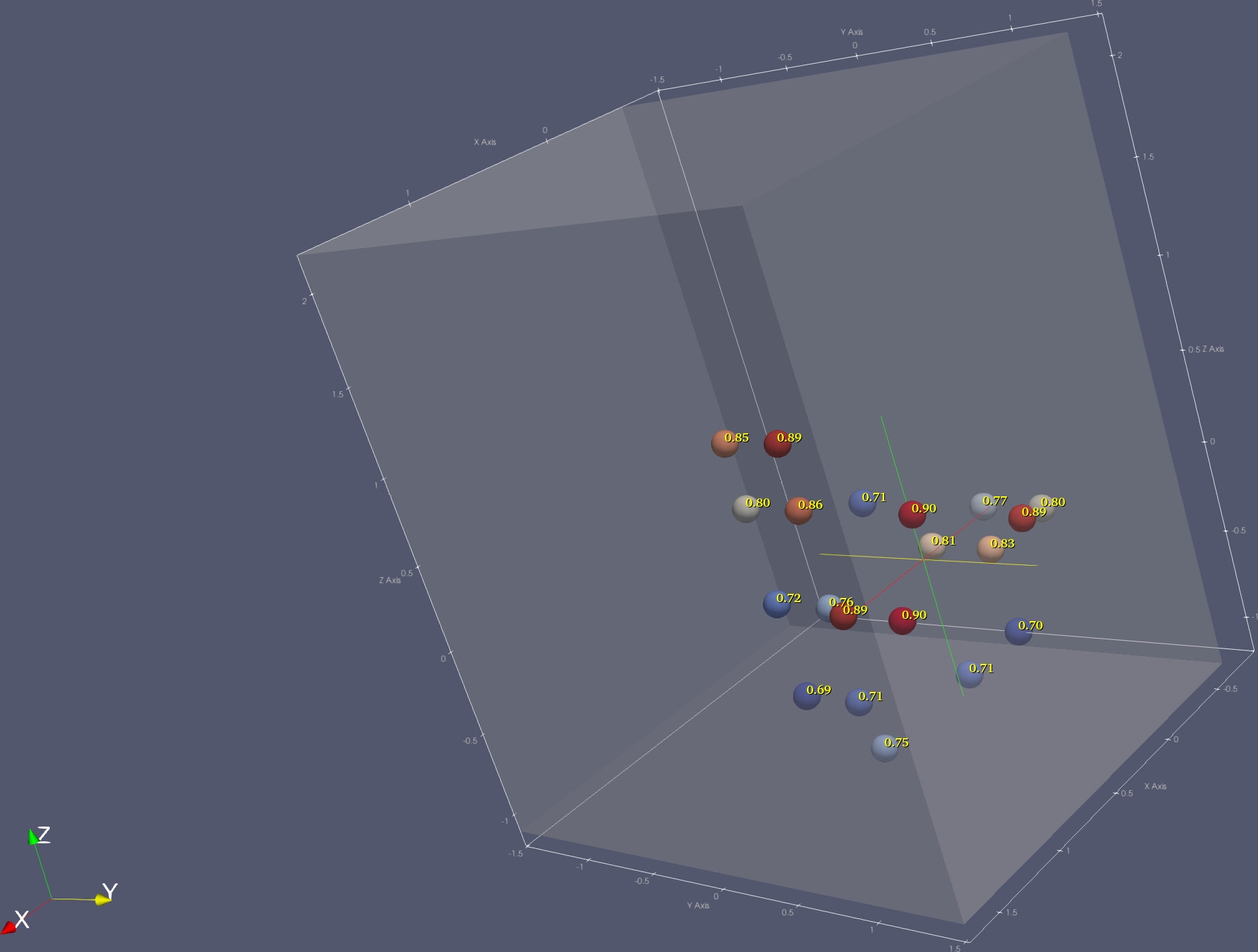}}

\bigskip

\stackinset{l}{0mm}{t}{1mm}{{\color{white}\textbf{\large (b)}}}{\includegraphics[width = 0.9\textwidth]{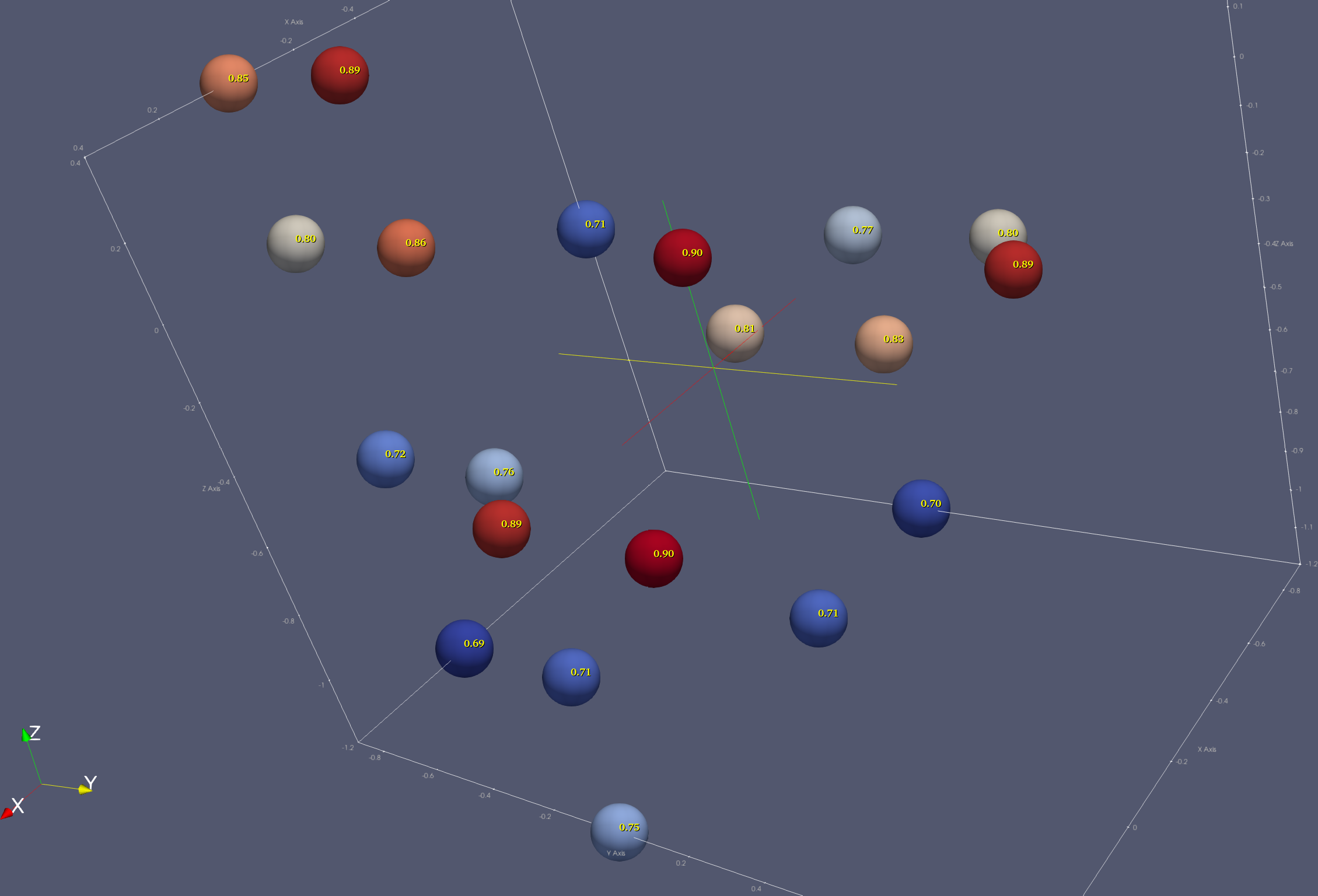}}
\caption{Location based apoptosis fractions with $U_{0}=1.5$ $\mu\text{M}$ based on 20 injection points. The numbers
attached to the points are the predicted apoptosis fractions at these locations.  \textbf{(b)} is a magnified
version of \textbf{(a)}.}
\label{fig:multiple_dose_center}
\end{figure}

We can further investigate this idea in a slightly different context.  Let us manually choose 20 different injection
points around the tumor region and calculate the apoptosis fractions by fixing the initial injection to $U_{0}=1.5$
$\mu\text{M}$.  We then label corresponding locations with these fractions and display them inside the tumor region
$\mathbb{T}$ in Fig. \ref{fig:multiple_dose_center}.  In the figure, blue points represent lower efficacy (69\%) and
red points represent higher efficacy (90\%) for an exposure of 72 hours.  It is observed that the injection location
has a significant influence on the efficacy of the drug.  For example, in Fig. \ref{fig:multiple_dose_center}, $P_4(0,1,0)$
(the right most red point) from Fig. \ref{fig:4CenterRes} is represented and had a much higher efficacy than several
points taken around the center.  In a realistic treatment case, one can increase the number of points used in the
simulations to obtain broader information about the optimal injection sites.  However, if the same figure is created
with a sufficient number of injection points around (and even outside) the tumor region, we can create a more
fine-grained apoptosis heat map.  In this way, we can obtain a volumetric partition of the tissue of interest with
respect to mean apoptosis fractions.  Such a work-flow can allow practitioners to determine the optimal infusion
locations. Once the corresponding partition is identified, we can utilize the simulation to find an ideal injection
amount which can strike a balance between toxicity and efficacy. 

In 3D simulations, computation time is undoubtedly of great importance and essentially determines if the
proposed model is feasible in practical applications. In finite element models, the type and number of elements
in the computational domain is the primary factor effecting the computation time. In this paper, we constructed
the domain with $98403$ bi-quadratic hexagonal elements. Our computing environment is 9th Generation
Intel(R) Core(TM) i7-9750H (12MB Cache, up to 4.5GHz, 6 cores) with 32GB DDR4-2666MHz RAM. We run
the simulations with an MPI-based parallel environment on four cores and observed that computation time
is approximately 15 minutes for one simulation with an initial injection of $U_{0}=1.5$ $\mu\text{M}$ and
an output of apoptosis fractions for a given threshold, $u_T(\tau)$, the results of which are reported in
Table \ref{tab:aptable} and 5.2 hours to create the dose-response curve using 31 initial injection values in
Fig. \ref{fig:dose_res}.  We should carefully note that the number of elements in this framework should be
completely determined by the dimension of the diffusion tensor volume extracted from the original data.
Thus, to be able to describe the corresponding quantities in finer detail, diffusion tensor images with higher
resolutions are needed, but in this case we need a more powerful environment to obtain the results in a
reasonable computing time.

\begin{figure}[htbp]
\centering  
\includegraphics[width=0.9\textwidth]{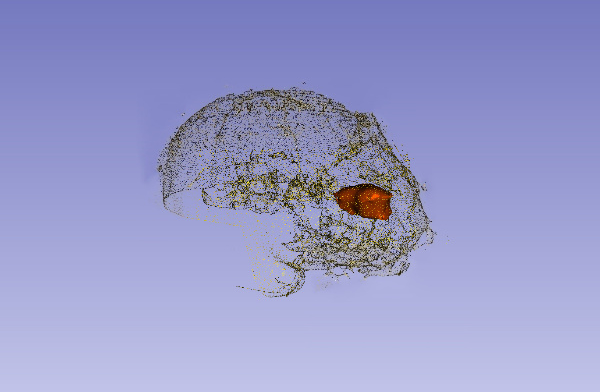}
\caption{Approximate tumor region segmented from the original DTI volume.  The solid tumor can be seen in red along
with a discretized representation of the brain}
\label{fig:3dtumor}
\end{figure}

In future studies, a more realistic computational domain may be created.
For simplicity, we worked with a cubic domain in this study.  However, it may possible to locate and cut out the tumor
region more accurately and create a volumetric mesh based on this segmented region.  A rough description of the tumor
region extracted from the original data can be seen in red in Fig. \ref{fig:3dtumor}.  We display the tumor along with a
discretized representation of the brain.  We can then feed this mesh structure into the existing model.  This approach
has the potential to generate a more realistic model, but it can be quite challenging. For example, we would need to
properly address how to interpolate the diffusion tensors across the boundary of this new domain if boundary conditions
are to be imposed only on the surface of the tumor volume.

\section{Conclusion and future work}
\label{Sec: Conclusion}

Brain tissue poses a unique transport challenge due to the highly inhomogeneous - anisotropic nature of the medium.
Since drug exposure directly impacts cell death, the geometry and topography of the tumor will have a significant
effect on efficacy.  Further, the tumor will also grow in an inhomogeneous - anisotropic manner
\cite{KCMSWMDA10, MLHKWAG11, CGFBAC15, EKS16}, and hence its structure can be quite unpredictable.  Diffusion
Tensor Magnetic Resonance Imaging (DTI) provides structure level information on an individual basis \cite{ODonnell-Westin2011}.
While there have been articles on employing DTI to study drug transport in the brain \cite{de2007drug}, thus far
none have simulated the efficacy of a drug as a consequence of drug transport.  With a drug transport - tumor
population coupled model, there is potential in producing computer aided treatment strategies.

In this paper, we derived an inhomogeneous - anisotropic drug diffusion model of molarity $u(\x,t)$, with the drug
mixture injected into a porous tumor region in Sec. \ref{Sec: GenericDiffusion}.  Then in Sec. \ref{Sec: Binary Population},
we developed the binary population model, where natural tumor cell death is equivalent to the rate of tumor population
growth and cell death due to the drug occurs when $u(\x,t) > u_T(\tau)$ for a concentration threshold $u_T$ and
exposure time $\tau$, similar to that of \cite{RGP18}.  The model is solved numerically, in Sec. \ref{Sec: Numerics}
through our finite element method.  Since information on the diffusion tensor is imperative to the numerical solution,
the model employs DTI data (Sec. \ref{Sec: DTI}) of a 35-year old male diagnosed with glioblastoma multiform (GBM).
This DTI data was pre-processed with an open source medical image processing software \textit{3DSlicer} \cite{3dslicer}
and fed into our finite element framework.  In Sec. \ref{Sec: Results} we present the results from the simulation in a
form that can be used by oncologists and doctors, and make predictions that may be easily overlooked in practice.
We investigated apoptosis fractions in the tumor region through the binary population model in Sec. \ref{Sec: Binary Population}
This allowed us to plot dose-response curves and test various injection locations.  Importantly, the various locations
show that the intuitive choice may not always be the best.  This opens the door to develop tools to aid oncologists
and doctors in deciding on an optimal treatment strategy.

The response of cancer cells to a therapeutic agent is undoubtedly a highly complex phenomenon. However, some aspects
of it can be addressed with the help of a mathematical model. In this sense, our aim in this study was to build a partial
differential equation framework based on patient-specific data that can be used to predict the efficacy of drug diffusion
in the brain tissue occupied by tumor cells.  Moreover, this framework may be used to create diffusion models which take
into account more complex considerations.  For example, a problem encountered in some cancer therapies is drug
resistance, which can be defined as the ability of cancer cells to survive and grow despite various anti-cancer
treatments \cite{leary2018sensitization}.  Further, oxygen concentration also has an effect on drug efficacy
\cite{Pappas2016, GAP16, KHP14, IRP13}.  In this paper, we assumed that the tumor cells exposed to a drug
concentration above certain threshold values will be ablated after the corresponding exposure times. However, if reliable
empirical knowledge is present about drug resistance and oxygen concentration, a time dependent or location-based
threshold model may be integrated into this framework. In this regard, using finite element method in the model provides a
great flexibility since it allows us to attach scalar quantities in the desired locations.

In addition, while the binary population model has computational advances due to its simplicity, a more accurate
model would be a stochastic dynamical system that is dependent on the concentration threshold.  One concern
may be the complexity of coupling, but fortunately transport happens on a much faster timescale than apoptosis.
We would solve the partial differential equations for the transport, and then tackle the population dynamics to
produce the apoptosis fractions.  However, as with any model, adding more complexity is counterproductive unless
it is accompanied by reliable data.  This gives us a singular opportunity to develop both biological and physical
experiments to accurately estimate parameters and test the models.
         
\section*{Acknowledgment}
E.K. and E.A. are grateful to NSF (grant \# DMS-1912902) for partial support of their contributions to this investigation.
E.K., A.R., E.A., and S.G. appreciate the support of the Department of Mathematics and Statistics at TTU, and 
S.G. also appreciates the support of the Department of Mathematics at UNL.
\bibliographystyle{unsrt}
\bibliography{Tumor_Paper_AminEdits}

\end{document}